\documentclass[aps,nopacs,nofootinbib,floatfix,superscriptaddress,onecolumn]{revtex4}
\usepackage{graphicx}
\usepackage{amsfonts}
\usepackage{amssymb}
\usepackage{amsbsy}
\usepackage{hyperref}
\usepackage{comment}
\usepackage{amsmath}
\usepackage{mathrsfs}
\usepackage{latexsym}
\usepackage{natbib}
\usepackage{bm}
\usepackage{subfigure} 
\usepackage{color}
\usepackage{wasysym}
\usepackage{mathbbol}
\usepackage{bigints}
\allowdisplaybreaks
\usepackage[normalem]{ulem}
\usepackage[dvipsnames]{xcolor}
\usepackage{multirow}

\definecolor{napiergreen}{rgb}{0.16, 0.5, 0.0}

\renewcommand*{\l}{\lambda_{\star}}

\def\nn{\nonumber} 

\def\f{\frac}
\def\l{\left}
\def\r{\right}
\def\d{{\rm d}}
\def\tx{\tilde{x}}

\def\cE{\mathcal{E}}

\def\erf{\mathcal{E}\text{rf}}
\DeclareMathOperator{\sech}{sech}

\def\erf{\mathcal{E}\text{rf}}
\def\erfi{\mathcal{E}\text{rfi}}
\def\erfc{\mathcal{E}\text{rfc}}

\begin{document}

\title{Signatures of gravitational wave memory in the radiative process of entangled quantum probes}

\author{Subhajit Barman}
\email{subhajit.barman@physics.iitm.ac.in}
\affiliation{Centre for Strings, Gravitation and Cosmology, Department of Physics, Indian Institute of Technology Madras, Chennai 600036, India}

\author{Indranil Chakraborty}
\email{indranil.phy@iitb.ac.in}
\affiliation{Department of Physics, Indian Institute of Technology Bombay, Mumbai 400076, India}

\author{Sajal Mukherjee}
\email{sajal.mukherjee@pilani.bits-pilani.ac.in} 
\affiliation{Department of Physics, Birla Institute of Technology and Science - Pilani, Rajasthan 333031, India}

\begin{abstract}
\noindent In this article, we examine entangled quantum probes in geodesic trajectories in a flat background with a gravitational wave (GW) burst. In particular, these quantum probes are prepared initially either in the symmetric or anti-symmetric Bell's states, and we study the radiative process as the GW burst passes. 
  {We split a generic GW burst into two profiles with and without memory.} GW burst with (without) memory profiles have different (similar) asymptotic strains between early and late times. We observe that for eternal switching, there is a finite change in the collective atomic transition rate due to the memory part of the GW burst, while the contribution from the without memory counterpart vanishes.
We also consider finite Gaussian switching and observe characteristic differences in the radiative process between the GW backgrounds with and without memory. Notably, if the Gaussian switching is peaked much later compared to the passing of GW, only the memory part contributes to the radiative process.
Thus, although examined in a simplified set-up, our findings suggest the potential to distinguish bursts with and without GW memory based on the radiative process of entangled detectors.

\end{abstract}

\date{\today}

\maketitle
%%%%%%%%%%%%%%%%%%%%%%%%%%%%%%%%%%%%%%%%%%%%%%%%%%%%%%%%%%%%%%%%%%%%%%%%%%%%%%
\section{Introduction}\label{sec:Introduction}
%%%%%%%%%%%%%%%%%%%%%%%%%%%%%%%%%%%%%%%%%%%%%%%%%%%%%%%%%%%%%%%%%%%%%%%%%%%%%%

The utilization of quantum field theory in curved spacetime results in outcomes that were unforeseen in classical general relativity (GR). One of the most celebrated of such outcomes is the Hawking effect, which predicts a thermal distribution of particles emitted from a black hole event horizon, as observed by an outside observer, see \cite{hawking1975}. Another subsequent interesting development is in the form of the Unruh effect \cite{Unruh:1976db}, where the Minkowski vacuum is seen populated by a thermal distribution of particles when observed from an accelerated frame. These effects are often regarded as a gateway to peek into the quantum features in a GR background \cite{book:Birrell, Don_Page:2004}. Over the years, there have been several efforts to detect these effects, see \cite{Martin-Martinez:2010, Aspachs:2010, Rideout:2012, Stargen:2021} (for an exhaustive list, see \cite{Crispino:2007eb} and the references therein). One of the most dedicated approaches in this line of investigation has been in the domain of analogue gravity, see  \cite{Rodriguez-Laguna:2016, Blencowe:2020, Gooding:2020, Biermann:2020, Onoe:2021, Hu:2018_nature}. However, as analogue gravity models lack the interpretation of curvature, they are often unable to provide a complete understanding of the intricacies related to a general curved background.\vspace{0.1cm}

Another promising avenue to investigate such features of semiclassical gravity has been in the field of relativistic quantum information (RQI), see \cite{Reznik:2002fz, Floreanini:2004, FuentesSchuller:2004xp, Ball:2005xa, Cliche:2009fma, Lin:2010zzb, MartinMartinez:2012sg, Salton:2014jaa, Martin-Martinez:2015qwa, Zhou:2017axh, Cai:2018xuo, Pan:2020tzf, Barman:2021oum, K:2023oon, Hsiang:2024qou} for its different aspects. Several set-ups in the domain of RQI involve employing entangled quantum systems for studying the entanglement due to motion \cite{Koga:2018the, Koga:2019fqh, Zhang:2020xvo, Barman:2022xht}, background geometries \cite{FuentesSchuller:2004xp, Cliche:2010fi, MartinMartinez:2012sg, Kukita:2017etu, Barman:2021kwg, Barman:2023rhd, Xu:2020pbj, Gray:2021dfk, K:2023oon}, thermal bath \cite{Brown:2013kia, Simidzija:2018ddw, Barman:2021bbw}, etc. A relatively under-explored though interesting area of research in RQI has been the study of {\em radiative process of entangled detectors}. Inspired by the light-matter interaction in atomic systems as introduced by Dicke \cite{Dicke:1954}, these works study the dynamics of entangled detectors and their dependence on various system parameters. Earlier works on the radiative process with static entangled detectors in flat spacetime have shown that the collective transitions are significantly different compared to single detector case and can get enhanced or reduced depending on the initial entangled state \cite{Arias:2015moa, FICEK2002369, Zhou:2020scl, Chen:2023xbc}. At the same time, entangled detectors in uniform acceleration and in contact with a thermal bath exhibit anti-Unruh-like effects, see  \cite{Barman:2021oum}. The radiative process of entangled detectors is also studied in other configurations, such as for detectors in circular trajectories \cite{Picanco:2020api, Barman:2022utm}, and in different black hole and cosmological spacetimes \cite{Menezes:2017oeb, Cai:2017jan, Liu:2018zod}.\vspace{0.1cm}

In this work, we consider a system comprising gravitational wave (GW) bursts passing through a flat background and investigate the radiative process of entangled detectors that are in geodesic trajectories. In recent times, the investigation of entanglement in gravitational wave backgrounds has received significant interest \cite{Xu:2020pbj, Gray:2021dfk, Chiou:2016exd, Barman:2023aqk, Wu:2024qhd}. For instance, the authors in \cite{Barman:2023aqk} have studied how spacetimes having GW bursts with and without memory control the concurrence measure of the harvested entanglement. GW memory refers to the permanent shift in the position of test particles upon the passage of a GW pulse \cite{Favata:2010,Chakraborty_EPJP:2022,Chakraborty:2021}. The change in the geodesic separation ($\Delta \xi^i$) is related to the change in GW metric perturbation ($\Delta h_{ij}$) as: 
\begin{equation}\label{eq:memory-exprsn}
 \Delta \xi^i=\frac{1}{2} \Delta h^i\,_j \xi^j~.
\end{equation}
This non-zero permanent change indicates the GW memory effect, see \cite{Tolish:2014bka, maggiore2007gravitational}. In the case of a burst with (without) memory, the asymptotic GW strain at past and future times are different (same) \cite{Barman:2023aqk}. Phenomenological models considering bursts with memory can be found in gravitational bremsstrahlung in hyperbolic orbits \cite{Kovacs:1978,Garcia-Bellido:2017,Hait:2022}, core-collapse supernova \cite{Mukhopadhyay:2021,Hait:2022}, and gamma-ray bursts \cite{Sago:2004}.   {In our present work, we split a GW burst into two profiles with and without memory propagating over Minkowski spacetime to investigate the radiative process.} The detectors in this scenario are considered to be the two-level atomic Unruh-DeWitt detectors. Our aim is to understand how GW memory controls the transitions between the two collective energy levels of entangled detectors. It is to be noted that all previous works investigating entanglement in a GW background have worked with static detectors. {In the present article, we consider detectors in geodesic trajectories that are initially prepared in an entangled state, either in the symmetric or in the anti-symmetric Bell states. However, the trajectories are so chosen that the separation between them is constant throughout the passage of the GW burst.}\vspace{0.1cm}

In this work, we consider both the eternal and finite switching between the detectors and the scalar field, where the finite switching is controlled by a Gaussian function. Both the eternal and the finite switching have their own significance. For instance, eternal switching are free of any transient effects and can bring out features that are specific to the background spacetime or the detector motion. On the other hand, finite switching is relevant from the experimental point of view, as one cannot construct a practical experimental set-up with detectors that interact for an infinite time with the background field. 
In particular, from our analysis, the key observation is that only the memory part of the GW burst ($\Delta h^i_j\neq 0$) gives a non-vanishing contribution in the radiative process for eternal switching. On the other hand, with finite switching, both the memory and non-memory parts of the GW burst can make a nonzero contribution to the radiative process. However, with finite switching, the memory part of the GW burst will always provide finite contribution if the detectors are switched on during or after the passing of the GW, while the non-memory part ($\Delta h^i_j=0$) will give non-zero finite contribution {\em only during the passing of the GW}, and the contribution vanishes much after the GW has passed.
We discuss the physical implications of these findings and emphasize the future prospects for experimental realizations.\vspace{0.1cm}

The manuscript is organized in the following manner. In Sec. \ref{sec:setup}, we briefly discuss the model set-up for studying the radiative process of entangled Unruh-DeWitt detectors. In this section, we also introduce the GW burst background and elucidate the detector trajectories, which are geodesics in the considered background. In Sec. \ref{eq:greens-fn}, we recall the expressions of the scalar field mode solutions in a GW background and estimate the expressions of the Green's function corresponding to the considered detector trajectories. Subsequently, in Sec. \ref{sec:Ke1-Rjl}, we utilize Green's functions to obtain different transition coefficients corresponding to the radiative process and also estimate the total transition probability for specific transitions in the infinite interaction (eternal switching) scenario. In section \ref{sec:KeG-Rjl} we consider the Gaussian switching scenario and find the transition coefficients and the total transition probabilities. We conclude this work with Sec. \ref{sec:discussion}, where we present our key observations and discuss the implications of our findings.

%%%%%%%%%%%%%%%%%%%%%%%%%%%%%%%%%%%%%%%%%%%%%%%%%%%%%%%%%%%%%%%%%%%%%%%%%%%%%%
\section{Set-up and detector trajectories}\label{sec:setup}
%%%%%%%%%%%%%%%%%%%%%%%%%%%%%%%%%%%%%%%%%%%%%%%%%%%%%%%%%%%%%%%%%%%%%%%%%%%%%%
In this section, we outline the model setup for the radiative process of two entangled quantum probes. We also discern the geodesic trajectories to be followed by these probes in the GW burst background. We shall observe that the collective transition of the entangled quantum probes not only depends on the individual detector Wightman functions but also on the Wightman function corresponding to the geodesic distance between two different detectors. Therefore, it is natural to believe that the effect of GW memory in these trajectories and the subsequent distortions in it should also be reflected in the collective transition of the entangled probes.

\subsection{Model set-up}

We start with a brief discussion on the set-up for the radiative process of entangled quantum probes. In this regard, we consider two entangled detectors in a background spacetime interacting with a massless, minimally coupled scalar field~$\Phi$. { The set-up is inspired by the works on light-matter interaction by Dicke (see ~\cite{Dicke:1954} for the original discussion and 
 ~\cite{Rodriguez-Camargo:2016fbq} for a recent one). In recent times, the set-up has been used in various scenarios and for different detector motions, see ~\cite{Arias:2015moa, Rodriguez-Camargo:2016fbq, Costa:2020aqa, Barman:2021oum, Barman:2022utm}. In particular, we consider the entangled quantum probes to be modeled after two-level, point-like, atomic detectors, popularly known as the Unruh-DeWitt detectors. These detectors were initially conceptualized to understand the Unruh and the Hawking effects (see  \cite{Unruh:1976db, hawking1975} on the conceptions of these effects), but in recent times have become ubiquitous in understanding entanglement phenomenon in curved spacetime, see  \cite{Xu:2020pbj, Barman:2021oum, Barman:2022utm, Barman:2023aqk}. Let us provide an outline for the set-up in which the complete system Hamiltonian is of the form
\begin{equation}\label{eq:Hamiltonian-total}
H = H_\mathrm{D}+H_\mathrm{F}+H_\mathrm{I}\,,
\end{equation}
where $H_\mathrm{D}$ denotes the free detector Hamiltonian, $H_\mathrm{F}$ is the free scalar field Hamiltonian, and $H_\mathrm{I}$ corresponds to the interaction between the detectors and the scalar field. The free detector Hamiltonian describes two static atoms, see 
~\cite{Dicke:1954}, can be expressed as
\begin{equation}
H_\mathrm{D} 
= \omega_{0}\, \l[\hat{S}_{1}^{z}\otimes\hat{\mathbb{1}}_{2}\,
+\, \hat{\mathbb{1}}_{1}\otimes \hat{S}_{2}^{z}\,\r]\,.\label{eq:HA}
\end{equation}
In this expression $\hat{S}_{j}^z$, with $j=\{1,2\}$, is the energy operator for the detectors, and it is defined as
\begin{equation}
\hat{S}_{j}^{z}
=\f{1}{2}\,\l(\vert e_{j}\rangle\, \langle e_{j}\vert
-\vert g_{j}\rangle\, \langle g_{j}\vert\r),
\end{equation}
where $\vert g_{j}\rangle$ and $\vert e_{j}\rangle$ respectively denote the ground and excited states of the $j$-$\mathrm{th}$ atomic detector. In Eq.~\eqref{eq:HA},~$\hat{\mathbb{1}}$ denotes the identity operator and $\omega_{0}$ the energy gap between different energy levels of the collective two detector system. As pointed out in ~\cite{Arias:2015moa}, for two identical and static detectors, the energy eigenstates and eigenvalues for the collective detector system can be expressed as
\begin{subequations}\label{eq:enrgy-lvls}
\begin{eqnarray}
E_{e} &=& \omega_{0},~~~~\vert e\rangle = \vert e_{1} \rangle\,
\vert e_{2}\rangle,\quad\\
E_{s} &=& 0,~~~~~~\vert s\rangle = \f{1}{\sqrt{2}}\, \l(\vert e_{1}\rangle\, 
\vert g_{2}\rangle + \vert g_{1}\rangle\, \vert e_{2}\rangle\r),\\
E_{a} &=& 0,~~~~~~\vert a\rangle = \f{1}{\sqrt{2}}\,\l(\vert e_{1}\rangle\, 
\vert g_{2}\rangle - \vert g_{1}\rangle\, \vert e_{2}\rangle\r),\quad\\
E_{g} &=& -\omega_{0},~~\vert g\rangle = \vert g_{1}\rangle \vert g_{2}\rangle,
\end{eqnarray}
\end{subequations}
where $\vert g\rangle$ and $\vert e\rangle$ denote the ground and 
the excited states of the collective system. On the other hand, $\vert s\rangle$ and $\vert a \rangle$ are the degenerate maximally entangled Bell states, also known as the symmetric and anti-symmetric Bell states. We have provided a pictorial representation of these states, energy levels, and the possible transitions between them in Fig.~\ref{fig:Energy-levels}.
%%%%%%%%%%%%%%%%%%%%%%%%%%%%%%%%%%%%%%%%%%%%%%%%%%%%%%%%%%%%%%%%%%%%%%%%%%%%%%%
\begin{figure}[!t]
\centering
\includegraphics[width=8.5cm]{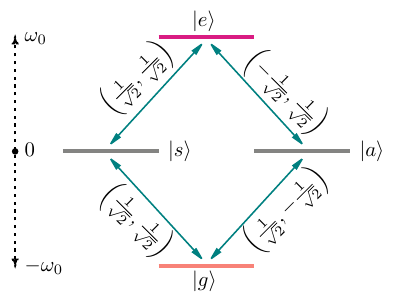}
\caption{The above illustration depicts the energy levels corresponding to the collective eigenstates of the two entangled Unruh-DeWitt detectors. Each of the detectors describes a two-level system. In the above figure, we have also indicated the expectation values of the individual detector monopole moment operators for each transition, where the first and second values, respectively, correspond to the two detectors.}\label{fig:Energy-levels}
\end{figure}
%%%%%%%%%%%%%%%%%%%%%%%%%%%%%%%%%%%%%%%%%%%%%%%%%%%%%%%%%%%%%%%%%%%%%%%%%%%%%%%

The interaction between the detectors and the field is assumed to be of monopole type, and the interaction Hamiltonian is explicitly expressed as 
\begin{equation}\label{eq:Hamiltonian-int}
H_\mathrm{I}
= \sum_{j=1}^{2}\mu\, m_j(\tau_{j})\,\kappa_{j}(\tau_{j})\,
\Phi[\tx_j(\tau_{j})],
\end{equation}
where we have considered the interaction strength $\mu$ to be the same for both the detectors. While $m_{j}(\tau_{j})$ and $\kappa_{j}(\tau_{j})$ respectively denote the monopole operators and the switching function corresponding to the interaction. In particular, for $j^{th}$ detector the monopole moment operator is given by
\begin{equation}
\hat{m}_{j}(0)=\vert e_{j}\rangle \langle g_{j}\vert
+ \vert g_{j}\rangle \langle e_{j}\vert.\label{eq:mo}
\end{equation} 
In the interaction picture, utilizing the previously mentioned interaction Hamiltonian the time evolution operator can be expressed as 
\begin{eqnarray}\label{eq:TimeEvolution-int}
\hat{U} &=& \mathcal{T}\; \exp\,\biggl\{-i\,\mu \sum_{j=1}^{2}\int_{-\infty}^{\infty}\,
\hat{m}_j(\tau_{j})\, \kappa_{j}(\tau_{j})\,
\hat{\Phi}[\tx_{j}(\tau_{j})]\,\d\tau_{j}\biggr\},
\end{eqnarray}
where $\mathcal{T}$ signifies the consideration of time ordering. Moreover, we consider the initial and final collective detector states to be $\vert \omega\rangle$ and $\vert \Omega\rangle$ respectively. Then treating the interaction strength $\mu$ perturbatively, one obtains the transition probability 
\begin{eqnarray}\label{eq:Transition-prob}
\Gamma_{\vert \omega\rangle\to\vert\Omega\rangle}(\cE) 
&\simeq & \mu^2 \sum_{j,l=1}^{2} 
m_{j}^{\Omega\omega\ast}\, 
m_{l}^{\Omega\omega}\,F_{jl}(\cE),
\end{eqnarray}
where $\cE=E_{\Omega}-E_{\omega}$, is the energy gap between the states $\vert \omega\rangle$ and $\vert \Omega\rangle$ with energies $E_{\omega}$ and $E_{\Omega}$. While $m_{j}^{\Omega\omega} 
= \langle \Omega \vert \hat{m}_{j}(0)\vert\omega\rangle$ denotes the expectation value of the monopole moment operator. One can use the expression of the monopole moment operator from Eq. (\ref{eq:mo}) to obtain these expectation values between different collective detector states. For instance, one can easily obtain for states~$\vert g \rangle$ and~$\vert e\rangle$ the expectation values $m_{j}^{ge} = m_{j}^{eg} =0$. Therefore, these transitions are not permissible. On the other hand, 
the other expectation values are $m_{1}^{se} = m_2^{se} =1/\sqrt{2}$ and $m_{1}^{ae} = -m_{2}^{ae} = -1/\sqrt{2}$. Similarly, with the Bell states and the ground state, the expectation values are $m_{1}^{gs} =m_{2}^{gs} =1/\sqrt{2}$ and $m_{1}^{ga}=-m_{2}^{ga}=1/\sqrt{2}$. These expectation values of the monopole moment operators are also depicted
in Fig.~\ref{fig:Energy-levels}.

In Eq. \eqref{eq:Transition-prob}, the most important part that contains the information about the background and the detector trajectories are $F_{jl}(\cE)$. Following  \cite{Barman:2022utm}, we refer to different components of $F_{jl}(\cE)$ as the auto or the cross-transition probabilities, depending on whether that component corresponds to a single detector or correlates two different ones. The explicit form of this quantity is given by
\begin{eqnarray}\label{eq:Transition-coeff}
F_{jl}(\cE) 
&=& \int_{-\infty}^{\infty}\d\tau_l'  \int_{-\infty}^{\infty}\d\tau_j\, 
\mathrm{e}^{-i\,\cE\,(\tau_{j}-\tau_{l}')}\nn\\
& &\times\, G_{jl}^{+}[\tx_{j}(\tau_j),\tx_{l}(\tau_l')]\, 
\kappa_j(\tau_j)\,\kappa_l(\tau_l'),
%\f{\d\tau_ {j}}{\d\tau}\, \f{d\tau_{l}}{\d\tau'},\nn\\
\end{eqnarray}
where the $G_{jl}^{+}[\tx_{j}(\tau_j),\tx_{l}(\tau_l')]$ denotes the 
positive frequency Wightman function evaluated along the detector trajectories and is defined as 
\begin{equation}\label{eq:Two-point-fn-gen}
G_{jl}^{+}[\tx_{j}(\tau_j),\tx_{l}(\tau_l')] 
= \langle 0 \vert \hat{\Phi}[\tx_{j}(\tau_{j})]\,
\hat{\Phi}[\tx_{l}(\tau_{l}')]\vert 0 \rangle,
\end{equation}
where $\vert 0 \rangle$ denotes a suitable field vacuum state. Our main aim in the subsequent sections is to evaluate these Wightman functions for detectors in geodesic trajectories in a GW background and then use them in discerning the transition probabilities of Eq. \eqref{eq:Transition-prob} for transitions from the symmetric and anti-symmetric Bell's states to the collective ground/excited state.

In our main analysis, we are going to consider two types of switching, namely the eternal switching with $\kappa(\tau)=1$ and finite switching in terms of Gaussian function $\kappa(\tau)=e^{-(\tau-\tau_{0})^2/2\sigma^2}$. In those scenarios, we define the transition probability rate from Eq. \eqref{eq:Transition-prob} as
\begin{eqnarray}\label{eq:Trans-rate-1}
    \mathcal{R}_{\omega\Omega} (\cE) &=& \frac{\Gamma_{\vert \omega\rangle\to\vert\Omega\rangle}(\cE)}{\mu^2\times\tilde{T}}~,
\end{eqnarray}
where $\tilde{T}$ signifies the effective duration for which the detectors are switched on. For eternal switching of $\kappa(\tau)=1$, and for the Gaussian switching of $\kappa(\tau)=e^{-(\tau-\tau_{0})^2/2\sigma^2}$ the expressions of $\tilde{T}$ are different. This $\tilde{T}$ is supposed to be obtained from the individual detector transition probability with the help of a change of variables $\eta=\tau_{j}-\tau'_{j}$ and $\xi=\tau_{j}+\tau'_{j}$. One can observe that for a time-translation invariant Green's function $G^{+}_{jj}\equiv G^{+}_{jj}(\tau_{j}-\tau'_{j})$, which is usually the case for detectors in inertial and uniformly accelerated trajectories, and for the considered switching the integral in the transition coefficient $F_{jj}(\cE)$ (when $j=l$) from Eq. \eqref{eq:Transition-coeff} can be decomposed into a multiplication of two parts. One entirely depends on $\eta$, which also contains the contribution from Green's function. Another one depends on $\xi$, does not contain any contribution from the Green's function, and may depend only on the switching function. The integration over this second term is often interpreted as the duration of interaction between the detectors and the field $\tilde{T}$, see  \cite{Barman:2022utm}. In particular, this integration over $\xi$ gives
\begin{widetext}
\begin{subequations}\label{eq:Trans-rate-2}
\begin{eqnarray}
    \tilde{T} &=& \lim_{T\to \infty}\,\int_{-T}^{T}d\xi~,~~ \textup{for}~\kappa(\tau_{j})=1~;\\
    \tilde{T} &=& \int_{-\infty}^{\infty}e^{-(\xi-2\tau_{0})^2/4\sigma^2}\,d\xi=2\sigma\sqrt{\pi}~,~~ \textup{for}~\kappa(\tau_{j}) = e^{-(\tau_{j}-\tau_{0})^2/2\sigma^2}~.
\end{eqnarray}
\end{subequations}
\end{widetext}
In Eq. \eqref{eq:Trans-rate-1} we have divided the transition probability by $\mu^2$ to define the transition rate. $\mu^2$ signifies the strength of interaction between the detectors and the background field that depends on the specific detector model. As it depends neither on the background spacetime nor on the detector motion, it is only natural to define the transition rate without this quantity.

\subsection{Trajectories for the quantum probes}

In this part of the section, we will find out the geodesic trajectories for the quantum probes in a gravitational wave background. In particular, the GW is assumed to be propagating in a flat background. As mentioned earlier, the geodesic trajectories in a flat spacetime are expected to get altered as the GW passes, which can have a direct relation with the memory effect. It is to be noted that in ~\cite{Barman:2023aqk} the existence of GW memory is recognized in entanglement with static detectors and the difference arises due to different background metrics (change in the metric $h_{ij}$ at early and late times). However, compared to ~\cite{Barman:2023aqk}, we believe our present work also captures the effect of GW memory by incorporating the effects of the geodesic motion of the detectors.  Moreover, we aim to understand the relationship between GW memory and different entanglement measures involving quantum probes.

We consider gravitational wave perturbations in a flat spacetime given by the line element (see ~\cite{Garriga-1991, Xu:2020pbj, Barman:2023aqk}) 
\begin{eqnarray}\label{eq:BJR-metric}
ds^2 &=& -du\,dv+dx^2\,[1+h(u)]+dy^2\,[1-h(u)]\,,
\end{eqnarray}
where we have considered the gravitational wave propagating along the $ z$-direction and $u=t-z$ and $v=t+z$. In the above expression, $h(u)$ denotes the specific profile of the GW burst. For instance, it can take forms of asymmetric functions such as the Heaviside theta and in terms of $\tanh$ functions, or it takes the forms of symmetric bursts such as the Gaussian and $\sech$-squared functions. However, for the time being, we are not going to put an explicit form of the burst profile for estimating the geodesic trajectories. In the background specified in Eq. \eqref{eq:BJR-metric}, the geodesic trajectories are obtained from the equations
\begin{eqnarray}\label{eq:Dtj-Minkowski-1}
    dU^{u}/d\tau &=& 0~,\nonumber\\
    dU^{x}/d\tau &=& -U^{u}U^{x}h'(u)~,\nonumber\\
    dU^{y}/d\tau &=& U^{u}U^{y}h'(u)~,\nonumber\\
    dU^{v}/d\tau &=& [(U^y){^2}-(U^x){^2}]h'(u)~;
\end{eqnarray}
where $h'(u)$ denotes the derivative of $h(u)$ with respect to $u$. From these equations, one can easily find out the velocity components as:
\begin{eqnarray}\label{eq:Dtj-Minkowski-2}
    U^{u} &=& C_{u}~,\nonumber\\
    U^{x} &=& C_{x}\,e^{-h(u)}~,\nonumber\\
    U^{y} &=& C_{y}\,e^{h(u)}~,\nonumber\\
    U^{v} &=& C_{v}+\frac{1}{2\,C_{u}}\,\Big[C_{y}^2\,e^{2h(u)}+C_{x}^2\,e^{-2h(u)}\Big]~;
\end{eqnarray}
where $C_{u}$, $C_{x}$, $C_{y}$, and $C_{v}$ are integration constants. One can integrate again the velocities $U^{\mu} = dx^{\mu}/d\tau$ to find out the coordinates $x^{\mu}$ in the geodesic trajectories. We find these trajectories to be 
\begin{widetext}
\begin{eqnarray}\label{eq:Dtj-Minkowski-3}
    u &=& C_{u}\,\tau+\tilde{C}_{u}~,\nonumber\\
    x &=& \frac{C_{x}}{C_{u}}\,\int e^{-h(u)}\,du + \tilde{C}_{x}~\approx \frac{C_{x}}{C_{u}}\, \big[u-\mathcal{A}\, \bar{g}(u)\big] + \tilde{C}_{x}~,\nonumber\\
    y &=& \frac{C_{y}}{C_{u}}\,\int e^{h(u)}\, du + \tilde{C}_{y}~ \approx \frac{C_{y}}{C_{u}}\, \big[u+\mathcal{A}\, \bar{g}(u)\big] + \tilde{C}_{y}~,\nonumber\\
    v &=& \frac{C_{v}}{C_{u}}\,u+\frac{1}{2\,C_{u}^2}\,\Big[C_{y}^2\,\int e^{2h(u)} du +C_{x}^2\,\int e^{-2h(u)} du\Big]+\tilde{C}_{v}~\nonumber\\
    &\approx& \frac{C_{v}}{C_{u}}\,u+\frac{1}{2\,C_{u}^2}\,\Big[C_{y}^2\,\big[u+2\mathcal{A}\, \bar{g}(u)\big] +C_{x}^2\,\big[u-2\mathcal{A}\, \bar{g}(u)\big]\Big]+\tilde{C}_{v}~.
\end{eqnarray}
\end{widetext}
Here $\{\tilde{C}_{\mu}\}$ is another set of constants of integration. In the previous expression, we have considered 
\begin{eqnarray}\label{eq:fu-gu-rel}
    h(u)=\mathcal{A}\,g(u)~,~~\mathrm{and}~~\Bar{g}(u)=\int g(u)\,du~;
\end{eqnarray}
where $\mathcal{A}$ denotes the strength of GW perturbation and $g(u)$ correspond to the specific burst profile.   {We would like to mention that this GW strength parameter $\mathcal{A}$ is linearly proportional to the Newton's constant $G$, see \cite{Jokela:2022rhk}.}

We consider some simple but sufficiently generic timelike geodesic trajectories which are obtained from the consideration of the parameter values $C_{u}=1$, $\tilde{C}_{u}=\tilde{C}_{v}=\tilde{C}_{y}=0$, $C_{x}=0$, $C_{y}=b$, and $C_{v}=c$. We consider the particular parameter $\tilde{C}_{x}=0$ for detector $A$ and $\tilde{C}_{x}=d$ for detector $B$. Then the geodesic trajectories are given by 
\begin{eqnarray}\label{eq:geodesic-trajectory}
    u_{j} &=& \tau_{j}~,\nonumber\\
    x_{j} &=&  d\,\delta_{j,B}~,\nonumber\\
    y_{j} &=& b\, \big[\tau_{j}+\mathcal{A}\, \bar{g}(\tau_{j})\big] ~,\nonumber\\
    v_{j} &=& c\,\tau_{j}+\frac{1}{2}\,\Big[ b^2\,\big\{\tau_{j}+2\mathcal{A}\, \bar{g}(\tau_{j})\big\}\Big]~.
\end{eqnarray}
In the above expressions, $\delta_{j,l}$ signifies the Kronecker delta. For these trajectories the condition of timelike trajectory $g_{\mu\nu}U^{\mu}U^{\nu}=-1$ results in the constraint
\begin{eqnarray}\label{eq:timelike-constraint}
    c-\frac{b^2}{2}=1~.
\end{eqnarray}
We shall utilize the expressions of the geodesic trajectories from Eq. \eqref{eq:geodesic-trajectory} along with the constraint \eqref{eq:timelike-constraint} to obtain the necessary Green's functions in the next section.
\color{black}

Note that in Eq.(\ref{eq:geodesic-trajectory}), the 
geodesic solutions are dependent on the GW amplitude and profile. Thus, the passage of a burst with memory will cause a permanent change in the positions of the detector. In an earlier work \cite{Barman:2023aqk}, the authors studied similar spacetime metric (given by Eq.(\ref{eq:BJR-metric})) with static Unruh-DeWitt detectors. In this article, we focus on geodesic detectors traversing the spacetime. Thus, the radiative process studied here captures both the effects of spacetime geometry and the detector motion.

%%%%%%%%%%%%%%%%%%%%%%%%%%%%%%%%%%%%%%%%%%%%%%%%%%%%%%%%%%%%%%%%%%%%%%%%%%%%%%
\section{Green's functions for detectors in geodesic trajectories}\label{eq:greens-fn}

We consider a massless minimally coupled scalar field $\Phi(\tilde{x})$ in the GW burst background specified by the line element (\ref{eq:BJR-metric}), where $\tilde{x}$ denotes specific detector trajectories. The scalar field equation of motion $\Box \Phi(\tilde{x}) = (1/\sqrt{-g})\,\partial_{\mu} \big[\sqrt{-g}\,g^{\mu\nu}\,\partial_{\nu}\Phi\big]=0$, in the background of Eq. (\ref{eq:BJR-metric}) takes the form 
\begin{eqnarray}\label{eq:scalar-field-EOM-1}
    -2\,\partial_{u}\partial_{v}\Phi +\frac{1}{2}\bigg[\frac{\partial_{x}^2}{1+h(u)} + \frac{\partial_{y}^2}{1-h(u)}\bigg]\,\Phi = 0~.
\end{eqnarray}
This equation accounts for mode solutions of the form of $\Phi \sim \mathcal{R}(u)\times\exp{\{i(-k_{-}v+k_{1}x+k_{2}y)\}}$, see ~\cite{Garriga-1991, Xu:2020pbj, Barman:2023aqk}. The normalized mode solution is given by 
\begin{widetext}
{ 
\begin{eqnarray}\label{eq:modeFn-general}
u_{\mathbf{k}}(\tilde{x}) &\simeq& \frac{1}{\sqrt{2k_{-}(2\pi)^3}}\,  \,\,\,\,\,\,\, \underbrace{\exp{\bigg[\frac{i\,\mathcal{A}}{4k_{-}}(k_{1}^2-k_{2}^2)\,\Bar{g}(u)\bigg]}}_{\text{background spacetime}}\,\,\,\,\,\, \,\, \underbrace{\exp \bigg[-i\,k_{-}v+i\,k_{1}x+i\,k_{2}y-i\frac{(k_{1}^2+k_{2}^2)}{4k_{-}}u \bigg] }_{\text{detector motion}}.
\end{eqnarray}}
\end{widetext}
Utilizing these mode functions, one can decompose the scalar field in terms of the ladder operators (~\cite{book:Birrell}) as
\begin{eqnarray}\label{eq:scalar-field-decomp}
\Phi(\tilde{x}) = \int\,d^3k\,\big[u_{\mathbf{k}}(\tilde{x})\,\hat{a}_{\mathbf{k}}+u^{\star}_{\mathbf{k}}(\tilde{x})\,\hat{a}^{\dagger}_{\mathbf{k}}\big]~.
\end{eqnarray}
Here the annihilation and the creation operators satisfy the commutation relation $[\hat{a}_{\mathbf{k}}, \hat{a}^{\dagger}_{\mathbf{k}'}] = (2\pi)^3\delta(\mathbf{k}-\mathbf{k}')$, and the operator $\hat{a}_{\mathbf{k}}$ annihilates the vacuum, say $|0\rangle$. Using this commutation relation and the mode expansion from Eq. (\ref{eq:scalar-field-decomp}), one can find out the Wightman function as $\langle 0|\,\Phi(\tilde{x})\,\Phi(\tilde{x}')\,|0\rangle = \int\,d\mathbf{k} \,u_{\mathbf{k}}(\tilde{x})\,u^{\star}_{\mathbf{k}}(\tilde{x}')$.

{From Eq. (\ref{eq:modeFn-general}), we observe that the mode solutions encapsulate two distinct effects: (a) the influence of the background geometry, and (b) the motion of the detector. This dual dependence is a hallmark of quantum systems in curved spacetime, as similarly seen in quantum entanglement setups (see \cite{Koga:2018the, Robbins:2020jca}). Hence, in the present context, both effects induce GW contributions in the Wightman function. However, in order to disentangle both the contributions and also to understand how our result reconciles with the Minkowski limit, we will expand the Eq. (\ref{eq:modeFn-general}) in two steps. In the following, let us briefly lay down the framework.  First, we will consider the change in background spacetime while remaining agnostic about the detector motion.  Next, we will incorporate the trajectory of the detector to get the final Wightman function.

{  To achieve the first step, we assume the GW perturbation amplitude $\thickmuskip=0mu \mathcal{A}\ll 1$.} The Wightman function obtained is,
\begin{eqnarray}\label{eq:Wightman-fn-general}
    G_{W}(\tilde{x},\tilde{x}') = G_{W}^{0}(\tilde{x},\tilde{x}') + G_{W}^{1}(\tilde{x},\tilde{x}')~.
\end{eqnarray}
As discussed earlier, here $G_{W}^{0}(\tilde{x},\tilde{x}')$ and $G_{W}^{1}(\tilde{x},\tilde{x}')$ are obtained by taking $\mathcal{O}(\mathcal{A}^{0})$ and $\mathcal{O}(\mathcal{A}^{1})$ terms from the series expansion of the
factor responsible for the background spacetime.  In particular, the general forms of $G_{W}^{0}(\tilde{x},\tilde{x}')$ and $G_{W}^{1}(\tilde{x},\tilde{x}')$ are given by
\begin{subequations}\label{eq:greensFn-MGW-general}
\begin{eqnarray}
    G_{W}^{0}(\tilde{x},\tilde{x}') &=& \frac{1}{4\pi^2\,\Delta u}\,\frac{1}{\frac{\sigma_{M}}{\Delta u}+i\,\epsilon}\\
    G_{W}^{1}(\tilde{x},\tilde{x}') &=& -\frac{\mathcal{A}(\Delta x^2-\Delta y^2)}{4\pi^2\,\Delta u^{3}}\,\frac{\left\{\Bar{g}(u)-\Bar{g}(u')\right\}}{\left(\frac{\sigma_{M}}{\Delta u}+i\,\epsilon\right)^2}~,
\end{eqnarray}
\end{subequations}
where, $\sigma_{M}\equiv -\Delta u\,\Delta v+\Delta x^2+\Delta y^2$. For detectors static at fixed spatial points, $G_{W}^{0}(\tilde{x},\tilde{x}')$ will indeed be independent of any GW perturbations, see  \cite{Barman:2023aqk}. This is basically the 
Wightman function computed in Minkowski spacetime.}

Now we utilize the expression of the trajectories from Eq. (\ref{eq:geodesic-trajectory}) and evaluate the necessary quantities to obtain these Green's functions. We have
\begin{widetext}
\begin{subequations}\label{eq:sigmaM-Dx2-exprsn}
\begin{eqnarray}
    \sigma_{M}(\tau_{1},\tau_{2}) &=& -\Delta \tau^2+\mathcal{A}\, \left\{\Bar{g}(\tau_{1})-\Bar{g}(\tau_{2})\right\}\,b^2\,\Delta \tau+d^2~\\
    \Delta x^2-\Delta y^2 &=& -b^2\,\Delta \tau^2-\mathcal{A}\, \left\{\Bar{g}(\tau_{1})-\Bar{g}(\tau_{2})\right\}\,2\,b^2\,\Delta \tau+d^2~,
\end{eqnarray}
\end{subequations}
where $\Delta\tau=\tau_{1}-\tau_{2}$. Then the components of the Wightman function specified in Eq. (\ref{eq:greensFn-MGW-general}) are obtained as\\
\begin{subequations}\label{eq:greensFn-MGW-general-2}
\begin{eqnarray}
    G_{W}^{0}(\tilde{x}_{1},\tilde{x}_{2}) &\simeq& -\frac{1}{4\pi^2}\, \frac{1}{\Delta \tau^2-d^2-i\,\epsilon\,\Delta \tau}-\frac{1}{4\pi^2}\, \frac{\mathcal{A}\, \left\{\Bar{g}(\tau_{1})-\Bar{g}(\tau_{2})\right\}\,b^2\,\Delta \tau}{(\Delta \tau^2-d^2-i\,\epsilon\,\Delta \tau)^2}~,\\
    G_{W}^{1}(\tilde{x}_{1},\tilde{x}_{2}) &\simeq& -\frac{1}{4\pi^2\,\Delta \tau}\, \frac{\mathcal{A}\, \left\{\Bar{g}(\tau_{1})-\Bar{g}(\tau_{2})\right\}\,\left[-b^2\,\Delta \tau^2 +d^2\right]}{(\Delta \tau^2-d^2-i\,\epsilon\,\Delta \tau)^2}~.
\end{eqnarray}
\end{subequations}
To arrive at these expressions, we have utilized the condition that $\mathcal{A}\ll 1$. One can add up the above two quantities to obtain the net expression of the Wightman function, which has a form of
\begin{eqnarray}\label{eq:greensFn-MGW-general-3}
    G_{W}(\tilde{x}_{1},\tilde{x}_{2}) &\simeq& -\frac{1}{4\pi^2}\, \frac{1}{\Delta \tau^2-d^2-i\,\epsilon\,\Delta \tau}-\frac{1}{4\pi^2}\, \frac{\mathcal{A}\, \left\{\Bar{g}(\tau_{1})-\Bar{g}(\tau_{2})\right\}}{(\Delta \tau^2-d^2-i\,\epsilon\,\Delta \tau)^2}\,\left[\frac{d^2}{\Delta \tau}\right]~.
\end{eqnarray}
\end{widetext}
Here, if one puts $d=0$, i.e., if one considers a single detector rather than taking two different detectors, the above Wightman function boils down to the form of 
\begin{eqnarray}\label{eq:WightmanFn-single-probe}
    G_{W}(\tilde{x}_{1},\tilde{x}_{2}) &\simeq& -\frac{1}{4\pi^2}\, \frac{1}{\Delta \tau^2-i\,\epsilon\,\Delta \tau}~,
\end{eqnarray}
which is the Minkowski spacetime Wightman function for a static detector. Therefore, even for a detector in a certain geodesic trajectory in a gravitational wave background, zero particle creation could be possible if eternal switching is considered. An interesting feature to note is that although Eq. (\ref{eq:greensFn-MGW-general-3}) was computed for geodesic detectors with constant separation, it remains identical to the scenario of two static detectors with the same constant separation up to $\mathcal{O}(\mathcal{A})$, see \cite{Xu:2020pbj, Barman:2023aqk}. In Appendix A, we demonstrate how this equivalence breaks down between static and geodesic detectors at $\mathcal{O}(\mathcal{A}^2)$.

In subsequent sections, we elaborate on the radiative process of two entangled detectors for the concerned background and motion using the Wightman functions obtained here. In this regard, we consider two specific scenarios. In one case, the detectors interact with the background scalar field for an infinite time (the eternal switching scenario), and in the other case, the detectors interact for a finite time with some suitable switching function.
\color{black}

%%%%%%%%%%%%%%%%%%%%%%%%%%%%%%%%%%%%%%%%%%%%%%%%%%%%%%%%%%%%%%%%%%%%%%%%%%%%%%

\section{Radiative process with eternal switching}\label{sec:Ke1-Rjl}
Here we consider the eternal switching, i.e., $\kappa(\tau_{j})=1$. We will be evaluating $F_{jj}$ and $F_{jl}$ step by step in the following chapters, and then finally estimate the total transition probability. In particular, we consider the initial detector states of being in the symmetric or anti-symmetric Bell states, i.e., either in $|s\rangle$ or in $|a\rangle$ (see Eq. \eqref{eq:enrgy-lvls}) and will investigate the radiative process for transitions to the collective ground or excited states.

\subsection{Evaluation of $F_{jj}$}

We consider the eternal switching scenario of $\kappa(\tau_{j})=1$ and proceed to evaluate $F_{jj}$. We consider a change of variables $\eta=\tau_{j}-\tau'_{j}$ and $\xi=\tau_{j}+\tau'_{j}$. Taking the expressions of the response functions from Eq. (\ref{eq:Transition-coeff}) and utilizing the Wightman function from Eq. (\ref{eq:WightmanFn-single-probe}) we obtain
\begin{eqnarray}\label{eq:Fjj-Ke1}
    F_{jj}(\mathcal{E}) &=& -\frac{1}{8\pi^2} \int^{\infty}_{-\infty}d\xi \int^{\infty}_{-\infty}d\eta \frac{e^{-i\,\mathcal{E}\,\eta}}{(\eta-i\,\epsilon/2)^2}~\nonumber\\
    &=& -\frac{\mathcal{E}\,\theta(-\mathcal{E})}{4\pi}\int^{\infty}_{-\infty}d\xi~.
\end{eqnarray}
One can obtain the expression of the transition probability rate $R_{jj}(\mathcal{E})$ for eternal switching by defining $R_{jj}(\mathcal{E})=F_{jj}/(\lim_{T\to \infty}\,\int_{-T}^{T}d\xi)$, as done in Eqs. \eqref{eq:Trans-rate-1} and \eqref{eq:Trans-rate-2}, which in this scenario provides us with the expression
\begin{eqnarray}\label{eq:Rjj-Ke1}
    R_{jj}(\mathcal{E}) 
    &=& -\frac{\mathcal{E}\,\theta(-\mathcal{E})}{4\pi}~.
\end{eqnarray}
It is to be noted that this quantity denotes the individual detector transition probability rate, and as observed from the above expression, is non-vanishing for $\mathcal{E}<0$, i.e., this transition probability rate is non-zero only for de-excitations from the entangled initial states.

\subsection{Evaluation of $F_{jl}$ with $j\neq l$}

We consider the general expression of $F_{jl}$ from Eq. (\ref{eq:Transition-coeff}) with the expression of the Wightman function from Eq. (\ref{eq:greensFn-MGW-general-3}). Here we consider a change of variables $\Bar{\eta}=\tau_{j}-\tau_{l}$ and $\Bar{\xi}=\tau_{j}+\tau_{l}$. One can write this expression as a sum of Minkowski and GW contributions, i.e., in terms of a contribution that is independent of $\mathcal{A}$ and another one that is dependent on $\mathcal{A}$. This sum looks like 
\begin{eqnarray}\label{eq:Fjl-full}
    F_{jl} = F_{jl}^{M}+ F_{jl}^{GW}~,
\end{eqnarray}
where $F_{jl}^{M}$ denotes the Minkowski part and $F_{jl}^{GW}$ denotes the GW part. We shall evaluate both of these quantities step by step. We begin by considering the eternal switching scenario $\kappa(\tau_{j})=1$ to evaluate these quantities. For instance, for eternal switching, the Minkowski part becomes
\begin{eqnarray}\label{eq:Fjl-M-Ke1}
    F_{jl}^{M} &=& -\frac{1}{8\pi^2} \int^{\infty}_{-\infty}d\Bar{\xi} \int^{\infty}_{-\infty}d\Bar{\eta} \frac{e^{-i\,\mathcal{E}\,\Bar{\eta}}}{(\Bar{\eta}-\frac{i\,\epsilon}{2}+d)(\Bar{\eta}-\frac{i\,\epsilon}{2}-d)}~\nonumber\\
    &=& -\frac{\sin{(\mathcal{E}\,d)}\,\theta(-\mathcal{E})}{4\pi\,d}\int^{\infty}_{-\infty}d\Bar{\xi}~.
\end{eqnarray}
Similar to the $F_{jj}$ terms, here also one can define the cross-transition rate as $R_{jl}^{M}=F_{jl}^{M}/(\lim_{T\to \infty}\,\int_{-T}^{T}d\xi)$, which turns out to be $R_{jl}^{M}=-\sin{(\mathcal{E}\,d)}\,\theta(-\mathcal{E})/(4\pi\,d)$. One can notice that in this scenario, $R_{jl}^{M}$ is non-vanishing only when $\mathcal{E}<0$. On the other hand, for eternal switching, the GW part is given by 
\begin{widetext}
\begin{eqnarray}\label{eq:Fjl-GW-Ke1-i}
    F_{jl}^{GW} &=& -\frac{\mathcal{A}}{8\pi^2} \int^{\infty}_{-\infty}d\Bar{\xi} \int^{\infty}_{-\infty}d\Bar{\eta} \frac{e^{-i\,\mathcal{E}\,\Bar{\eta}}\,\left\{\Bar{g}\left(\frac{\Bar{\xi}+\Bar{\eta}}{2}\right)-\Bar{g}\left(\frac{\Bar{\xi}-\Bar{\eta}}{2}\right)\right\}}{(\Bar{\eta}-\frac{i\,\epsilon}{2}+d)^2(\Bar{\eta}-\frac{i\,\epsilon}{2}-d)^2}\, \left[\frac{d^2}{\Bar{\eta}}\right]~\nonumber\\
    &=& -\frac{\mathcal{A}\,\theta(-\mathcal{E})}{16\pi\,d^2} \int^{\infty}_{-\infty}d\Bar{\xi} \,\Bigg[2\left\{\Bar{g}\left(\frac{\Bar{\xi}+d}{2}\right)-\Bar{g}\left(\frac{\Bar{\xi}-d}{2}\right)\right\}\times\left\{2\,\sin{(d\,\mathcal{E})}-\mathcal{E}\,d\,\cos{(d\,\mathcal{E})}\right\}\nonumber\\
    ~&& ~ +d\,\sin{(d\,\mathcal{E})}\times\left\{g\left(\frac{\Bar{\xi}+d}{2}\right)+g\left(\frac{\Bar{\xi}-d}{2}\right)\right\}\Bigg]\,.
\end{eqnarray}
\end{widetext}
This expression can be simplified to 
\begin{eqnarray}\label{eq:Fjl-GW-Ke1-f}
    F_{jl}^{GW} &=& -\frac{\mathcal{A}\,\theta(-\mathcal{E})}{8\pi\,d}\,\sin{(d\,\mathcal{E})} \int^{\infty}_{-\infty}d\Bar{\xi}~ g\left(\frac{\Bar{\xi}}{2}\right)~.
\end{eqnarray}
From the previous expression, one can notice that depending on different forms of the GW burst profiles $g(u)$, different cross-transition probabilities among the detectors can be obtained. 
We shall consider a generic burst profile and demonstrate how the memory part of it influences the qualitative features of $F_{jl}^{GW}$.\\

  {We split a general form of the GW burst into two profiles with and without memory}, specifically given by $h(u)=\alpha\,h_{S}(u)+\beta\,h_{A}(u)$. In this expression, $f_{S}(u)$ denotes a transient function of $u$, which vanishes at early and late times. At the same time, $f_{A}(u)$, corresponding to the memory part, has a difference in the asymptotic strain values at early and late times.   {The parameters $\alpha$ and $\beta$ are book-keeping parameters to track the contributions of the terms with and without memory.} In particular, we shall consider the values of $\alpha$ and $\beta$ between $0$ and $1$ depending on specific needs. Previously we mentioned that $h(u)=\mathcal{A}\,g(u)$, where $\mathcal{A}$ is the strength of the GW perturbation and $g(u)$ the specific functional profile of the GW burst. Thus, the general form of the GW burst profile is chosen to be,
\begin{eqnarray}\label{eq:gen-GW-profile}
    g(u)=\alpha\,g_{_{S}}(u)+\beta\,g_{_{A}}(u)~.
\end{eqnarray}
With the help of Eq. \eqref{eq:fu-gu-rel} we define $\Bar{g}(u)=\int g(u)\,du=\alpha\,\bar{g}_{_{S}}(u)+\beta\,\bar{g}_{_{A}}(u)$. Then with the previous expression of Eq. \eqref{eq:gen-GW-profile} we obtain the component of the cross-transition coefficient due to GW from Eq. \eqref{eq:Fjl-GW-Ke1-f} as
\begin{widetext}
\begin{eqnarray}\label{eq:Fjl-GW-Ke1-f2}
    F_{jl}^{GW} &=& -\frac{\mathcal{A}\,\theta(-\mathcal{E})}{4\pi\,d}\,\sin{(d\,\mathcal{E})} ~ \bigg[\alpha\,\bar{g}_{_{S}}\left(\frac{\Bar{\xi}}{2}\right)+\beta\,\bar{g}_{_{A}}\left(\frac{\Bar{\xi}}{2}\right)\bigg]^{\infty}_{\bar{\xi}\to -\infty}~.
\end{eqnarray}
\end{widetext}
Now, one can expect that for any function $g_{_{S}}(\bar{\xi}/2)$, vanishing at $\xi\to\pm\infty$, its integrated form $\bar{g}_{_{S}}(\bar{\xi}/2)$ should be finite in the asymptotic regions as $\bar{\xi}\to -\infty$ or $\bar{\xi}\to \infty$. On the contrary, for GW burst with memory profiles, the quantity $\bar{g}_{_{A}}(\bar{\xi}/2)$ can diverge out in the asymptotic limits as in one of the limits the integrand does not vanish. Let us understand this with a few specific examples. For instance, if one considers the symmetric profiles to be $g_{_{S}}(\bar{\xi}/2)=e^{-(\bar{\xi}/2)^2/\rho^2}$ or $g_{_{S}}(\bar{\xi}/2)= \sech^2{\{\bar{\xi}/(2\, \varrho)\}}$, their integration will give $\bar{g}_{_{S}}(\bar{\xi}/2)=(\rho\,\sqrt{\pi}/2)~\mathrm{Erf} \left(\bar{\xi}/2\rho\right)$ or $\bar{g}_{_{S}}(\bar{\xi}/2)=\varrho  \tanh \left(\bar{\xi}/\varrho\right)$. Both of these expressions of $\bar{g}_{_{S}}(\bar{\xi}/2)$ are finite in the asymptotic limit. Let us denote, in Eq. \eqref{eq:Fjl-GW-Ke1-f2}, the quantity $\big[\bar{g}_{_{S}} \left(\Bar{\xi} /2\right)\big]^{\infty}_{\bar{\xi}\to -\infty}=\Delta \bar{g}_{_{S}}$, which is finite.
On the other hand, let us consider $g_{_{A}}(\bar{\xi}/2)=\theta (u)$ or $g_{_{A}}(\bar{\xi}/2)= \{1+\tanh{(\bar{\xi}/2\lambda)}\}/2\,$ for bursts with memory profiles. The ensuing integration will give $\bar{g}_{_{A}}(\bar{\xi}/2)=(\bar{\xi}/2)\,\theta (\bar{\xi}/2)$ or $\bar{g}_{_{A}}(\bar{\xi}/2)=(\bar{\xi}/4)+(\lambda/2) \, \log \left[\cosh \left(\bar{\xi}/2\lambda\right)\right]$. Both of these integrated out expressions are divergent in the asymptotic limit and can be cast into a form to provide $\big[\bar{g}_{_{A}} \left(\Bar{\xi} /2\right) \big]^{\infty}_{\bar{\xi}\to -\infty}=(1/4)\int^{\infty}_{-\infty}d\Bar{\xi}$, in Eq. \eqref{eq:Fjl-GW-Ke1-f2}. Then we can simplify the expression of \eqref{eq:Fjl-GW-Ke1-f2} to provide
\begin{eqnarray}\label{eq:Fjl-GW-Ke1-f3}
    F_{jl}^{GW} &=& -\frac{\mathcal{A}\,\theta(-\mathcal{E})}{4\pi\,d}\,\sin{(d\,\mathcal{E})} ~ \bigg[\alpha\,\Delta \bar{g}_{_{S}}+\frac{\beta}{4}\,\int^{\infty}_{-\infty}d\Bar{\xi}\bigg]~.
\end{eqnarray}
With the help of Eqs. \eqref{eq:Trans-rate-1} and \eqref{eq:Trans-rate-2} we can find the cross-transition probability rate due to GW, which is defined as $R_{jl}^{GW}=F_{jl}^{GW}/(\lim_{T\to \infty}\,\int_{-T}^{T}d\Bar{\xi})$. In particular, considering the cross-transition probability from Eq. \eqref{eq:Fjl-GW-Ke1-f3} we obtain this rate to be
\begin{eqnarray}\label{eq:Rjl-GW-Ke1}
    R_{jl}^{GW} 
    &=& -\frac{\mathcal{A}\,\theta(-\mathcal{E})}{4\pi\,d}\,\l(\frac{\beta}{4}\r)\, \sin{(d\,\mathcal{E})}~.
\end{eqnarray}
We could arrive at the above expression since $\Delta \bar{g}_{_{S}}$ is finite for all GW profiles as $g_{_{S}}(u)$ vanishes at asymptotic limits. The above expression \eqref{eq:Rjl-GW-Ke1} asserts that the cross-transition probability rate due to GW depends only on the memory part. With the help of Eq. \eqref{eq:Rjl-GW-Ke1} one can easily find the collective transition probability rate for the atoms. In particular, we consider that these atoms are initially prepared in the symmetric or anti-symmetric Bell states ($|s\rangle$ and $|a\rangle$) and the final state is the collective ground or excited state $|g\rangle$ or $|e\rangle$. With the help of Eq. \eqref{eq:Transition-prob} and the expectation values of the monopole moment operators from Sec. \ref{sec:setup}, we define the collective transition probability rates, which can also be obtained from Eqs. \eqref{eq:Trans-rate-1} and \eqref{eq:Trans-rate-2}, as
\begin{widetext}
\begin{subequations}\label{eq:Rsg-ag-Gen-K1}
\begin{eqnarray}
    \mathcal{R}_{sg/se} &=& \frac{\Gamma_{\vert s\rangle\to\vert g\rangle/\vert e\rangle}}{\mu^2\,\l(\lim_{T\to \infty}{\int^{T}_{-T}d\xi}\r)}=\frac{1}{2}\l[R_{11}+R_{22}+(R_{12}+R_{21})\r]~,\\
    \mathcal{R}_{ag/ae} &=& \frac{\Gamma_{\vert a\rangle\to\vert g\rangle/\vert e\rangle}}{\mu^2\,\l(\lim_{T\to \infty}{\int^{T}_{-T}d\xi}\r)}=\frac{1}{2}\l[R_{11}+R_{22}-(R_{12}+R_{21})\r]~.
\end{eqnarray}
\end{subequations}
\end{widetext}
Here, the values of $R_{jl}$ are obtained from Eqs. \eqref{eq:Rjj-Ke1}, \eqref{eq:Fjl-M-Ke1}, and \eqref{eq:Rjl-GW-Ke1}. It is to be noted that in the above expressions, only the transitions from the symmetric and anti-symmetric Bell states to the collective ground state are possible. The reason is excitations ($\cE>0$) are not permissible with eternal switching, which is evident from Eqs. \eqref{eq:Rjj-Ke1}-\eqref{eq:Rjl-GW-Ke1}. We get this transition probability rate to the collective ground state as
\begin{subequations}\label{eq:Rsg-ag-Theta-K1}
\begin{eqnarray}\label{eq:Rsg-Theta-K1}
    \mathcal{R}_{sg} &=& -\frac{\theta(-\mathcal{E})}{4\pi}\l[\mathcal{E}+\frac{\sin{(\mathcal{E}\,d)}}{d}+\beta\,\frac{\mathcal{A}\,\sin{(d\,\mathcal{E})}}{4\,d}\r]~,\\
    \mathcal{R}_{ag} &=& -\frac{\theta(-\mathcal{E})}{4\pi}\l[\mathcal{E}-\frac{\sin{(\mathcal{E}\,d)}}{d}-\beta\,\frac{\mathcal{A}\,\sin{(d\,\mathcal{E})}}{4\,d}\r]~.\label{eq:Rag-Theta-K1}
\end{eqnarray}
\end{subequations}
In these expressions, by putting $\mathcal{A}=0$ one can obtain the Minkowski results as discussed in \cite{Arias:2015moa} (see Eqs. $2.16$ and $2.17$ of \cite{Arias:2015moa}). From these above expressions, one can also observe that $\mathcal{R}_{sg}-\mathcal{R}_{ag}=-\l[1+\beta\,\mathcal{A}/4\r]\,\{\theta(-\mathcal{E})\,\sin{(d\,\mathcal{E})}\}/(2\pi\,d)$, which only depends on the presence of memory terms in the GW profile, i.e., when $\beta\neq 0$. Therefore, the presence of GW memory modifies the difference between $\mathcal{R}_{sg}$ and $\mathcal{R}_{ag}$ by a factor of $\l[1+\beta\,\mathcal{A}/4\r]$ when compared to the purely Minkowski background.
Thus one can define a quantity $\Delta\mathcal{R}_{sg-ag} =(\mathcal{R}_{sg}-\mathcal{R}_{ag})\big|_{GWB}-(\mathcal{R}_{sg}-\mathcal{R}_{ag})\big|_{Mink}$, which is the difference in $(\mathcal{R}_{sg}-\mathcal{R}_{ag})$ between the background spacetimes containing GWs and without it.  It is evident that this quantity $\Delta\mathcal{R}_{sg-ag}$ is non-vanishing only when $\beta$ is non-zero, i.e., when the GW burst contains memory.

\color{black}

\section{Radiative process with finite Gaussian switching}\label{sec:KeG-Rjl}
Here we consider finite switching in terms of the Gaussian functions  $\kappa(\tau)=e^{-(\tau-\tau_{0})^2/2\sigma^2}$. Like the eternal switching scenario here also we shall evaluate $F_{jj}$ and $F_{jl}$ and estimate the total transition probability.

\subsection{Evaluation of $F_{jj}$}
To evaluate $F_{jj}$ we consider a change of variables $\eta=\tau_{j}-\tau'_{j}$ and $\xi=\tau_{j}+\tau'_{j}$. In terms of this change of variables, we get the quantity $\kappa(\tau_{j})\kappa(\tau'_{j})=\exp{\l[-\l\{(\xi-2\,\tau_{0})^2+\eta^2\r\}/4\sigma^2\r]}$. We utilize this expression and the expression of the Wightman function from Eq. (\ref{eq:WightmanFn-single-probe}) to obtain the response functions from Eq. (\ref{eq:Transition-coeff}) as
\begin{eqnarray}\label{eq:Fjj-KeG}
    F_{jj} &\simeq& -\frac{1}{8\pi^2} \int_{-\infty}^{\infty}d\xi\int_{-\infty}^{\infty}d\eta\, \frac{e^{-i\,\mathcal{E}\,\eta}}{(\eta-i\,\epsilon/2)^2}\,e^{-\l\{(\xi-2\,\tau_{0})^2+\eta^2\r\}/4\sigma^2}\nonumber\\
    ~&=& \frac{1}{4\pi}\l[e^{-\sigma^2\,\mathcal{E}^2}-\sqrt{\pi}\,\sigma\,\mathcal{E}\,\erfc(\sigma\,\mathcal{E})\r]~.
\end{eqnarray}
To obtain this expression we have used the Gaussian integral formula $\int_{-\infty}^{\infty}d\xi\,e^{-\alpha\,(\xi-\beta)^2}=\sqrt{\pi/\alpha}$, and the Fourier transform of the Gaussian integral $e^{-\eta^2/4\sigma^2}=(\sigma/\sqrt{\pi})\int_{-\infty}^{\infty}d\zeta\,e^{-\zeta^2\,\sigma^2+i\,\zeta\,\eta}$. One can find the explicit steps to evaluate this result in  \cite{Sriramkumar1996FinitetimeRO, Barman:2023aqk} and also in  \cite{Xu:2020pbj}. Here also one can define the transition probability rate from $F_{jj}$ by dividing it with $(2\,\sigma\sqrt{\pi})$ which is obtained from the integration $\int_{-\infty}^{\infty}d\xi \, e^{-(\xi-2\,\tau_{0})^2/4\sigma^2}=2\,\sigma \sqrt{\pi}$. The same transition probability rate for the Gaussian switching is also defined in Eqs. \eqref{eq:Trans-rate-1} and \eqref{eq:Trans-rate-2}. However, it is not always necessary as in this scenario, we are dealing with finite switching and can obtain a certain interaction time by fixing $\sigma$. 

\subsection{Evaluation of $F_{jl}$ with $j\neq l$}
To evaluate $F_{jl}$, we consider a change of variables $\Bar{\eta} = \tau_{j} - \tau_{l}$ and $\Bar{\xi}=\tau_{j}+\tau_{l}$, and in a manner similar to Eq. \eqref{eq:Fjl-full} express the cross-transition term as $F_{jl} = F_{jl}^{M}+ F_{jl}^{GW}$. With the help of Eqs. \eqref{eq:Transition-coeff} and \eqref{eq:greensFn-MGW-general-3} the Minkowski part of this cross-transition can be obtained as
\begin{widetext}
\begin{eqnarray}\label{eq:Fjl-M-KeG-1}
    F_{jl}^{M} &\simeq& -\frac{1}{8\pi^2} \int_{-\infty}^{\infty}d\Bar{\xi}\int_{-\infty}^{\infty}d\Bar{\eta}\, \frac{e^{-i\,\mathcal{E}\,\Bar{\eta}}\,e^{-\l\{(\Bar{\xi}-2\,\tau_{0})^2+\Bar{\eta}^2\r\}/4\sigma^2}}{\{(\Bar{\eta}-i\,\epsilon/2)^2-d^2\}}\nonumber\\
    ~&=& -\frac{\sigma^2}{8\pi^2d} \int_{-\infty}^{\infty}d\zeta\,e^{-\zeta^2\sigma^2}\int_{-\infty}^{\infty}d\Bar{\eta}~e^{i\,(\zeta-\mathcal{E})\,\Bar{\eta}}\l[\frac{1}{(\Bar{\eta}-d-i\,\epsilon/2)}-\frac{1}{(\Bar{\eta}+d-i\,\epsilon/2)}\r]~,
\end{eqnarray}
where we have used the Gaussian integration formula $\int_{-\infty}^{\infty}d\Bar{\xi}\,e^{-\alpha\,(\Bar{\xi}-\beta)^2}=\sqrt{\pi/\alpha}$ and the Fourier transformation of the Gaussian function $e^{-\Bar{\eta}^2/4\sigma^2}=(\sigma/\sqrt{\pi})\int_{-\infty}^{\infty}d\zeta\,e^{-\zeta^2\,\sigma^2+i\,\zeta\,\Bar{\eta}}$. Considering a contour in the upper half complex plane and using the Residue theorem one can evaluate the $\Bar{\eta}$ integration and obtain

\begin{eqnarray}\label{eq:Fjl-M-KeG-2}
    F_{jl}^{M} &\simeq& -\frac{i\,\sigma^2}{4\pi\,d} \int_{-\infty}^{\infty}d\zeta\,e^{-\zeta^2\sigma^2}\,\l[e^{i\,(\zeta-\mathcal{E})\,d}-e^{-i\,(\zeta-\mathcal{E})\,d}\r]\nonumber\\
    ~&=& \frac{\sqrt{\pi}\,\sigma\,e^{-d^2/4\sigma^2}}{4\pi\,d}\,\l[\sin{(-d\,\mathcal{E})}-\mathcal{R}e\l\{i\,e^{i\,d\,\mathcal{E}}\,\erf\l(\sigma\,\mathcal{E}+\frac{i\,d}{2\,\sigma}\r)\r\}\r]~,
\end{eqnarray}
where we have used the identities of the \emph{Error functions} $\erfc\,{(z)}=1-\erf\,{(z)}$ and $\erf\,{(-z)}=-\erf\,{(z)}$. From the above expression, it is easy to find out the rate of transition $R_{jl}^{M}(\sigma)=F_{jl}^{M}(\sigma)/(2\,\sigma\sqrt{\pi})$. In the limit of eternal switching $\sigma\to \infty$ the previous expression of $R_{jl}^{M}$ reduces to
\begin{eqnarray}\label{eq:Rjl-M-KeG-Ke1-limit}
    \lim_{\sigma\to \infty}\,R_{jl}^{M}(\sigma) = \frac{\theta(-\mathcal{E})\,\sin{(-d\,\mathcal{E})}}{4\pi\,d}~,
\end{eqnarray}
where we have used the asymptotic expansion of the error function $\erf\,{(z)}= 1-e^{-z^2}\l\{1/(\sqrt{\pi}\,z)+\mathcal{O}(1/z^3)\r\}$ for $z\to \infty$. One can notice that this expression is the same as provided in Eq. \eqref{eq:Fjl-M-Ke1}, corresponding to eternal switching.\\

On the other hand, with the help of the expressions from Eqs. \eqref{eq:Transition-coeff} and \eqref{eq:greensFn-MGW-general-3}, the purely GW part of the cross-transition $F_{jl}^{GW}$ can be obtained in the following manner
\begin{eqnarray}\label{eq:Fjl-GW-KeG-1}
    F_{jl}^{GW} &\simeq& -\frac{\mathcal{A}}{8\pi^2} \int_{-\infty}^{\infty}d\Bar{\xi}\int_{-\infty}^{\infty}d\Bar{\eta}\,e^{-i\,\mathcal{E}\,\Bar{\eta}}\,e^{-\l\{(\Bar{\xi}-2\,\tau_{0})^2+\Bar{\eta}^2\r\}/4\sigma^2}~\l(\frac{d^2}{\Bar{\eta}}\r)~\frac{\bar{g}\l(\frac{\Bar{\xi}+\Bar{\eta}}{2}\r)-\Bar{g}\l(\frac{\Bar{\xi}-\Bar{\eta}}{2}\r)}{\{(\Bar{\eta}-i\,\epsilon/2)^2-d^2\}^2}\,.
\end{eqnarray}
After carrying out the $\Bar{\eta}$ integration, which is done utilizing the Residue theorem with the help of the Fourier transform formula $e^{-\Bar{\eta}^2/4\sigma^2}=(\sigma/\sqrt{\pi})\int_{-\infty}^{\infty}d\zeta\,e^{-\zeta^2\,\sigma^2+i\,\zeta\,\Bar{\eta}}$, we obtain 
\begin{eqnarray}\label{eq:Fjl-GW-KeG-2}
    F_{jl}^{GW} &\simeq& -\frac{\mathcal{A}}{64\,\pi^{3/2}\,d^2\,\sigma ^2} \int_{-\infty}^{\infty}d\Bar{\xi}\,e^{-\frac{d^2+(\Bar{\xi} -2 \tau_{0})^2}{4 \sigma ^2}} \Bigg[\sqrt{\pi } \sigma ^2 \left\{2 \sin (d \mathcal{E})-\mathcal{R}e\left(e^{-i d \mathcal{E}} \erfi\left(\frac{d}{2 \sigma }+i \sigma  \mathcal{E}\right)\right)\right\}\nonumber\\
    ~&\times& \left\{d\,g\left(\frac{d+\Bar{\xi} }{2}\right)+d\,g\left(\frac{\Bar{\xi} -d}{2}\right)-4\, \Bar{g}\left(\frac{d+\Bar{\xi} }{2}\right)+4\, \Bar{g}\left(\frac{\Bar{\xi} -d}{2}\right)\right\} \nonumber\\
    ~&+& 4 \sqrt{\pi } \,d\, \sigma^2 \,\mathcal{E} \left\{\mathcal{R}e\l(e^{i\,d\,\mathcal{E}}\erf\left(\sigma  \mathcal{E}+\frac{id}{2 \sigma }\right)\r) -\cos{(d\,\mathcal{E})}\right\} \, \left\{\Bar{g}\left(\frac{\Bar{\xi} -d}{2}\right)-\Bar{g}\left(\frac{d+\Bar{\xi} }{2}\right)\right\}\nonumber\\
    ~&+& 2\,d^2 \left\{\sqrt{\pi } \,\mathcal{R}e\l(i\,e^{i\,d\,\mathcal{E}}\erf\left(\sigma  \mathcal{E}+\frac{id}{2 \sigma }\right)\r) +\sqrt{\pi }\,\sin{(d\,\mathcal{E})}+\frac{2 \sigma \,e^{-i\,d\,\mathcal{E}}}{d}\, e^{\frac{\left(d+2 i \sigma ^2 \mathcal{E}\right)^2}{4 \sigma ^2}}\right\}\nonumber\\
    ~&\times& \left\{\Bar{g}\left(\frac{\Bar{\xi} -d}{2}\right)-\Bar{g}\left(\frac{d+\Bar{\xi} }{2}\right)\right\}\Bigg].
\end{eqnarray}
\end{widetext}

We shall use the same chosen GW profile as given in Eq. \eqref{eq:gen-GW-profile}. However, unlike the eternal switching scenario, the integral in Eq. (\ref{eq:Fjl-GW-KeG-2}) cannot be solved analytically. Thus, to evaluate the integration over $\Bar{\xi}$ and clarify the roles of memory and non-memory profiles, we select specific burst profiles $g(u)$, to reinforce our argument. The explicit forms of $g_{_{S}}(u)$ chosen are Gaussian and $\sech$-squared, and the explicit forms of $g_{_{A}}(u)$ chosen are  the Heaviside-theta and $\tanh$ functions. In particular, we shall provide numerical plots for the rate of this cross-transition probability due to GW, which with the help of Eqs. \eqref{eq:Trans-rate-1} and \eqref{eq:Trans-rate-2} is defined as $R^{GW}_{jl} (\cE) = F^{GW}_{jl} (\cE)/(2\sigma\sqrt{\pi})$.

%%%%%%%%%%%%%%%%%%%%%%%%%%%%%%%%%%%%%%%%%%%%%%%%%%%%%%%%%%%%%%%%%%%%%%%%%%%%%%%%%%%%%%%%%%%%%%%%%
\begin{figure}[h!]
\centering
\includegraphics[width=8.5cm]{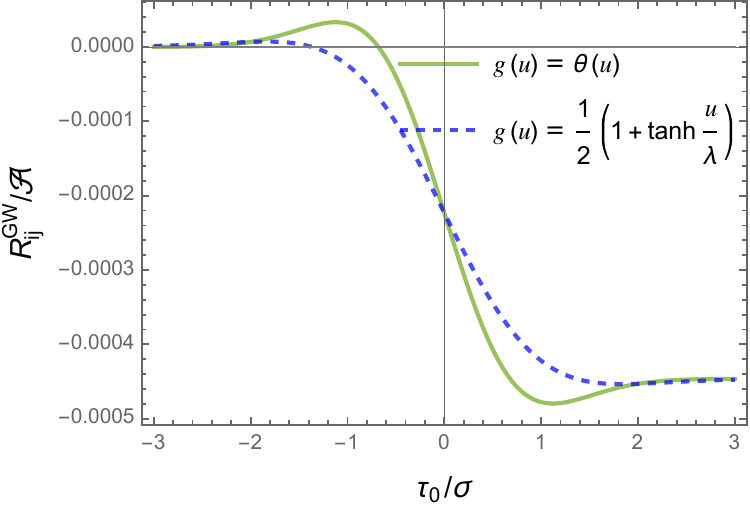}
\hskip 20pt
\includegraphics[width=8.5cm]{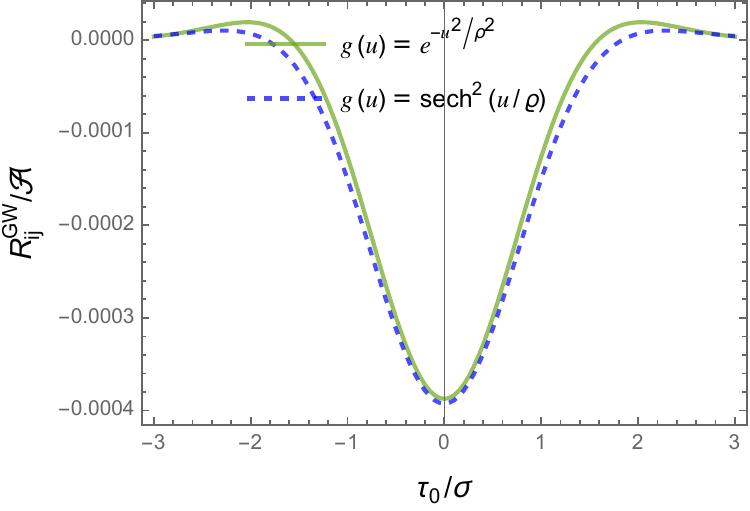}
\caption{  {\textbf{Left:} On the left, we have plotted the cross-transition probability rate due to GW $(R_{jl}^{GW})$ for a GW burst profile of \eqref{eq:gen-GW-profile} with $\alpha=0$ and $\beta=1$, i.e., only for bursts with memory. We have plotted $R_{jl}^{GW}$ for the GW burst profiles $g(u)=\theta(u)$ and $g(u)=\{1+\tanh{(u/\lambda)\}/2}$ as functions of the mean in switching time $\tau_{0}/\sigma$. To arrive at these plots, we have fixed the other parameters. \textbf{Right:} On the right, we have plotted $R_{jl}^{GW}$ with profile \eqref{eq:gen-GW-profile} when $\alpha=1$ and $\beta=0$, i.e., bursts without memory. In particular, we have plotted $R_{jl}^{GW}$ for the GW burst profiles $g(u)=e^{-u^2/\rho^2}$ and $g(u)=\sech^2{(u/\varrho)}$ as functions of the mean in switching time $\tau_{0}/\sigma$.   {When compared to the memory profiles, one can observe that the above curves for GW profiles without memory are symmetric with respect to $\tau_{0}$ around $\tau_{0}/\sigma=0$.}
Both of the above plots correspond to the Gaussian switching $\kappa(\tau)=e^{-(\tau-\tau_{0})^2/2\sigma^2}$.  The other parameters are fixed at the values: $d/\sigma=0.75$, $\sigma\,\mathcal{E}=0.5$, and $\rho/\sigma=0.75=\varrho/\sigma,$ $\lambda/\sigma=0.75$. }}
\label{fig:RijGW-vtau0}
\end{figure}
%%%%%%%%%%%%%%%%%%%%%%%%%%%%%%%%%%%%%%%%%%%%%%%%%%%%%%%%%%%%%%%%%%%%%%%%%%%%%%%%%%%%%%%%%%%%%%%%%

From the analysis concerning finite Gaussian switching functions, we observe that excitations from the Bell states to the collective excited state are possible. Moreover, all the GW burst profiles, with and without memory, contribute to the cross-transition probability. For GW bursts with memory, if the detectors are switched on after the passing of the GW one will always get a non-zero finite contribution in the GW-induced cross-transition probability rate $R_{jl}^{GW}$, see Fig. \ref{fig:RijGW-vtau0} for different profiles and Fig. \ref{fig:RijGW-vstau0-dalpha-beta} for a specific $g(u)$. At the same time, for bursts without memory, the contribution in the GW-induced cross-transition probability rate $R_{jl}^{GW}$ is maximum if the detectors are switched on during the passing of the GW, see Figs. \ref{fig:RijGW-vtau0} and \ref{fig:RijGW-vstau0-dalpha-beta}.   {To further our argument, we have chosen two distinct GW profiles, one with memory, and one without memory, and show the generic features of $R_{jl}^{GW}$}. We believe these observations can provide us with important insights regarding constructing experimental set-ups to identify GW memory. 
For instance, Fig. \ref{fig:RijGW-vstau0-dalpha-beta} highlights the significant impact of the memory profile on the behavior of $R_{jl}^{GW}$. In the left plot, we observe that if Gaussian switching is turned on much after the passing of the GW, the transition probability amplitude remains unchanged for non-memory profiles as long as the memory profile has the same amplitude. This observation is further supported by the right plot, where an increase in the memory profile amplitude leads to an increased transition probability rate at later times.

%%%%%%%%%%%%%%%%%%%%%%%%%%%%%%%%%%%%%%%%%%%%%%%%%%%%%%%%%%%%%%%%%%%%%%%%%%%%%%%%%%%%%%%%%%%%%%%%%
\begin{figure}[h!]
\centering
\includegraphics[width=8.5cm]{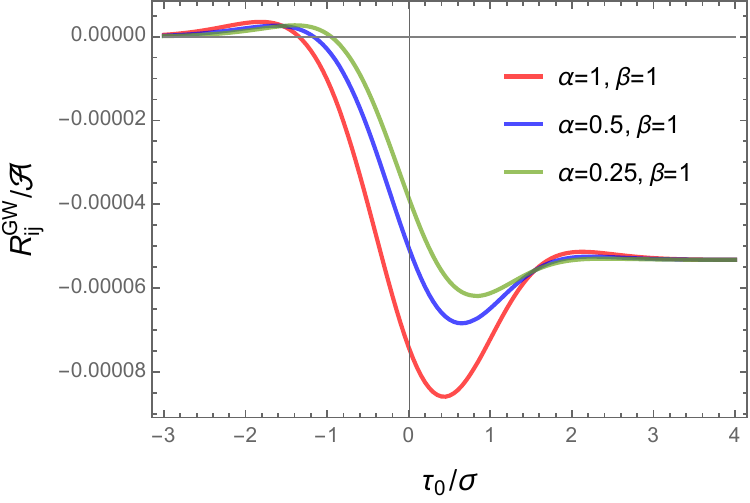}
\hskip 20pt
\includegraphics[width=8.5cm]{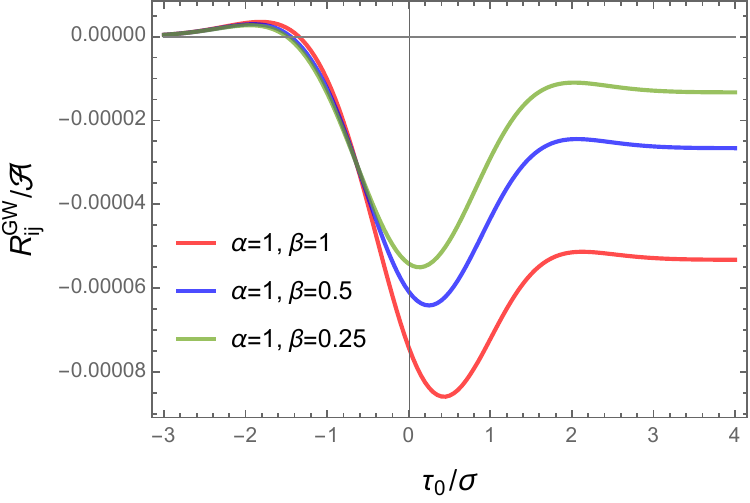}
\caption{  {\textbf{Left:} On the left, we have plotted the cross-transition probability rate due to GW $(R_{jl}^{GW})$ for a GW burst profile of \eqref{eq:gen-GW-profile} as functions of the mean in switching time $\tau_{0}/\sigma$. In this scenario, we have varying $\alpha$, $\beta=1$, $d/\sigma=0.5$, $\sigma\,\mathcal{E}=0.5$, and $\lambda/\sigma=0.75$. \textbf{Right:} On the right, we have plotted $R_{jl}^{GW}$ with GW profile \eqref{eq:gen-GW-profile} as functions of the mean in switching time $\tau_{0}/\sigma$. In this scenario, we have $\alpha=1$, varying $\beta$, $d/\sigma=0.75$, $\sigma\,\mathcal{E}=0.5$, and $\rho/\sigma=0.75=\varrho/\sigma$.
Both of the above plots correspond to the Gaussian switching $\kappa(\tau)=e^{-(\tau-\tau_{0})^2/2\sigma^2}$. The explicit form of the GW profile chosen is $g(u)=\alpha\,e^{-u^2/\rho^2}+ \beta\,\theta(u)$.}}
\label{fig:RijGW-vstau0-dalpha-beta}
\end{figure}
%%%%%%%%%%%%%%%%%%%%%%%%%%%%%%%%%%%%%%%%%%%%%%%%%%%%%%%%%%%%%%%%%%%%%%%%%%%%%%%%%%%%%%%%%%%%%%%%%

We shall now talk about the collective transition probability rate of the atoms when they interact with the background field via a Gaussian switching. Unlike the eternal switching scenario, the transition probabilities corresponding to Gaussian switching are finite, and both excitation and de-excitation are possible, see Eqs. \eqref{eq:Fjj-KeG}-\eqref{eq:Fjl-GW-KeG-2} and the Fig. \ref{fig:RijGW-vtau0}. In particular, we get $F_{jl}^{M}$ from Eq. \eqref{eq:Fjl-M-KeG-2} in analytical form and $F_{jl}^{GW}$ from Eq. \eqref{eq:Fjl-GW-KeG-2}  through the numerical integration.
With the Gaussian switching, the observed excitations are due to the transient nature of the switching and can lead to exciting outcomes, see  \cite{Sriramkumar1996FinitetimeRO}. Therefore, let us focus on these excitation probability rates. We mention that the de-excitation probability can be obtained from the excitation by considering a negative atomic energy gap. With the help of Eqs. \eqref{eq:Transition-prob} and \eqref{eq:Trans-rate-1}, we express the transition probability rates, from the symmetric or anti-symmetric Bell states to the collective excited state as
\begin{widetext}
\begin{subequations}\label{eq:Rsg-ag-Gen-KGauss}
\begin{eqnarray}
    \mathcal{R}_{se} &=& \frac{\Gamma_{\vert s\rangle\to\vert e\rangle}}{\mu^2\,(2\sigma\sqrt{\pi})}=\frac{1}{2}\l[R_{11}+R_{22}+(R_{12}+R_{21})\r]~,\\
    \mathcal{R}_{ae} &=& \frac{\Gamma_{\vert a\rangle\to\vert e\rangle}}{\mu^2\,(2\sigma\sqrt{\pi})}=\frac{1}{2}\l[R_{11}+R_{22}-(R_{12}+R_{21})\r]~.
\end{eqnarray}
\end{subequations}
From the previous discussion and the discussion in Sec. \ref{sec:KeG-Rjl}, we understood that $R_{jj}=R_{jj}^{M}$ and $R_{jl}=R_{jl}^{M}+R_{jl}^{GW}$ for $j\neq l$. Then from Eq. \ref{eq:Rsg-ag-Gen-KGauss} we can obtain the expressions $\mathcal{R}_{se}+\mathcal{R}_{ae} = R_{11}^{M}+R_{22}^{M}$  and $\mathcal{R}_{se}-\mathcal{R}_{ae} = R_{12}^{M} + R_{21}^{M} + \l(R_{12}^{GW} + R_{21}^{GW}\r)$.
One can notice that $(\mathcal{R}_{se}+\mathcal{R}_{ae})$ is free of any contribution from the GW. At the same time, similar to the infinite switching scenario, for finite Gaussian switching $(\mathcal{R}_{se}-\mathcal{R}_{ae})$ carries the effect of the GW burst. It is to be noted that if one subtracts out the Minkowski $(\mathcal{R}_{se}-\mathcal{R}_{ae})|_{Mink}$ from the above $(\mathcal{R}_{se}-\mathcal{R}_{ae})$, then the resulting quantity depends entirely on the GW part. Furthermore, because of the symmetry $R_{12}^{GW}=R_{21}^{GW}$, the difference will be exactly given by
\begin{eqnarray}\label{eq:DRsg-ag-Gen-KGauss-2}
    \Delta\mathcal{R}_{se-ae} &=&(\mathcal{R}_{se}-\mathcal{R}_{ae})\big|_{GWB}-(\mathcal{R}_{se}-\mathcal{R}_{ae})\big|_{Mink} = 2\,R_{12}^{GW}~,
\end{eqnarray}
\end{widetext}
where $(\mathcal{R}_{se}-\mathcal{R}_{ae})|_{GWB}$ denotes the relevant quantity estimated in the gravitational wave background. From Figs. \ref{fig:RijGW-vtau0} and \ref{fig:RijGW-vstau0-dalpha-beta}, we observe that each $R_{12}^{GW}$ from different GW burst backgrounds can have some signature reminiscent of the presence of GW memory. Thus one can notice that the quantity in Eq. \eqref{eq:DRsg-ag-Gen-KGauss-2} can be of help in isolating and identifying the GW memory as $R_{12}^{GW}=\Delta\mathcal{R}_{se-ae}/2$.

\color{black}

%%%%%%%%%%%%%%%%%%%%%%%%%%%%%%%%%%%%%%%%%%%%%%%%%%%%%%%%%%%%%%%%%%%%%%%%%%%%%%
\section{Discussion \& concluding remarks}\label{sec:discussion}
%%%%%%%%%%%%%%%%%%%%%%%%%%%%%%%%%%%%%%%%%%%%%%%%%%%%%%%%%%%%%%%%%%%%%%%%%%%%%%

In this article, we have investigated the radiative process of two entangled Unruh-DeWitt detectors initially prepared in the maximally entangled Bell states. The detectors follow geodesic trajectories in a spacetime having linearized GW burst propagating over a flat background.   {We have chosen to work with a generic GW burst profile \eqref{eq:gen-GW-profile}, having contributions from both the memory and non-memory terms.} Furthermore, we have considered two types of switching, namely eternal switching of $\kappa(\tau)=1$ and finite Gaussian switching of $\kappa(\tau)=e^{-(\tau-\tau_{0})^2/2\sigma^2}$, to investigate the features of the radiative process in the concerned background. The key observations from our investigation and their implications are as follows.

\begin{widetext}
\begin{table}[]
    \centering
    \begin{tabular}{ ||p{4.5cm}|p{3cm}|p{3cm}|p{6.0cm}||  }
 \hline
 \multicolumn{4}{|c|}{A tabular list of our observations} \\
 \hline
 Switching type& GW profile: $g(u)$ & Presence of memory & Features in the radiative process\\
 \hline
 \multirow{2}{*}{Eternal: $\kappa(\tau)=1$} & $\alpha=0$, $\beta=1$ & With memory & Non vanishing $R_{jl}^{GW}$ \\\cline{2-4}
 & $\alpha=1$, $\beta=0$ & Without memory & Vanishing $R_{jl}^{GW}$ \\
 \hline
 \multirow{2}{*}{Gaussian: $\kappa(\tau)=e^{-\frac{(\tau-\tau_{0})^2}{2\sigma^2}}$} & $\alpha=0$, $\beta=1$ & With memory & Non-zero $|R_{jl}^{GW}|$ in $\tau_{0}/\sigma\gtrsim 0$ \\\cline{2-4}
 & $\alpha=1$, $\beta=0$ & Without memory & Non-zero $|R_{jl}^{GW}|$ peaked around $\tau_{0}/\sigma=0$ \\
 \hline
\end{tabular}
    \caption{The above table lists our key observations regarding the radiative process of entangled Unruh-DeWitt detectors corresponding to different switching functions and GW burst profiles.}
    \label{tab:Obs}
\end{table}
\end{widetext}

\begin{itemize}
    \item For eternal switching, we find that the individual detector excitation in geodesic trajectories in a GW background vanishes, considering terms up to the first order in GW strength, i.e., taking terms $\mathcal{O}(\mathcal{A})$. This phenomenon, evident from Eq. \eqref{eq:Rjj-Ke1}, is akin to the static detector scenario (see  \cite{Barman:2023aqk}). Thus, it shows that the passage of a gravitational wave does not alter the particle content of a background, confirming the proposal in  \cite{Gibbons:1975}. It is to be noted that even with finite Gaussian switching, the individual detector transition probability does not contain any contribution from the GW (see Eq. \eqref{eq:Fjj-KeG}) and thus becomes identical with the one obtained in a Minkowski background (in this regard see  \cite{Sriramkumar1996FinitetimeRO}). It should be mentioned that non-vanishing particle creation for finite switching is attributed to the transient nature of the switching function, see  \cite{Sriramkumar1996FinitetimeRO, Barman:2022utm}.

    \item   {In regard to entangled detectors, we observed that for eternal switching no excitation is possible, and only de-excitations are possible, see Eqs. (\ref{eq:Rjj-Ke1}-\ref{eq:Fjl-GW-Ke1-f3}). This observation implies that with eternal switching, transitions from the Bell states to the collective excited state are not permitted, while transitions to the collective ground state are possible. On the other hand, with finite Gaussian switching, excitations, i.e., transitions to the collective excited state, are also possible, see Eqs. (\ref{eq:Fjj-KeG}-\ref{eq:Fjl-GW-KeG-2}). This observation signifies that it is imperative to consider finite switching, if one measures only the collective excitation of an entangled system that follows geodesic trajectories to report the passage of GW.}

    \item In regard to entangled detectors and eternal switching, we found that the GW-induced cross-transition probability rate is non-zero for the burst profiles with GW memory, and it vanishes for the GW bursts without memory. This fact is evident from the expression of $R_{jl}^{GW}$ from Eq. \eqref{eq:Rjl-GW-Ke1}. Furthermore, as observed in Eq. \ref{eq:Rsg-ag-Theta-K1} due to the presence of GW the difference in the total transition probability rates between the symmetric and anti-symmetric Bell states $(\mathcal{R}_{sg}-\mathcal{R}_{ag})$ will have a modified coefficient $\l(1+\mathcal{A}/2\r)$ for the bursts with memory as compared to unity corresponding to the bursts without memory. Thus, one could conclude that due to GW memory, there is a finite change in the total transition probability rates for eternal switching, which is absent for bursts without memory.
    We have observed that the quantity $\Delta\mathcal{R}_{sg-ag}$, which represents the difference in $(\mathcal{R}_{sg}-\mathcal{R}_{ag})$ between the gravitational wave and Minkowski backgrounds, directly captures the presence of GW memory.

    \item Regarding entangled detectors and Gaussian switching, we observed that the GW-induced cross-transition probabilities are non-zero for both GW bursts with and without memory. In particular, with only the burst without memory profile, we observed that $|R_{jl}^{GW}|$ is peaked around $\tau_{0}/\sigma=0$, i.e., when the Gaussian switching function peaks at the origin time of the GW, Fig. \ref{fig:RijGW-vtau0}. At the same time, with the memory profile, we found that it attains a saturated finite value after $\tau_{0}/\sigma=0$, and in the region $\tau_{0}/\sigma<0$, it vanishes, Fig. \ref{fig:RijGW-vtau0}. Note that with GW memory, in the region $\tau_{0}/\sigma>0$, $|R_{jl}^{GW}|$ is non-zero and finite. Thus, there will be a finite change in the transition probability for bursts with memory even after the GW has passed, which is confirmed in Fig. \ref{fig:RijGW-vstau0-dalpha-beta}. This observation can form a basis for a future experimentally viable test of GW memory.

   \item Examining $R_{jl}^{GW}$ for the joint scenario containing both bursts with and without memory, we find that the transition probability rate is unaffected by the burst amplitude in non-memory profiles if Gaussian switching is turned on much after the passing of the GW, as long as the amplitude of the memory profile remains unchanged. Moreover, in a similar scenario, increasing the amplitude of the memory profile leads to a change in the transition probability rate, see Fig. \ref{fig:RijGW-vstau0-dalpha-beta}.
    
    \item With the help of Eq. \eqref{eq:DRsg-ag-Gen-KGauss-2} we inferred that by measuring a quantity $\Delta\mathcal{R}_{se-ae}$, that signifies the difference in $(\mathcal{R}_{se}-\mathcal{R}_{ae})$ between the GW and Minkowski backgrounds, one can isolate the contribution of GW-induced cross-transition $R_{12}^{GW}$. In particular, this quantity is then given by $R_{12}^{GW}=\Delta\mathcal{R}_{se-ae}/2$. Then, in more simple terms, one can identify the effects of GW memory by measuring $\Delta\mathcal{R}_{se-ae}$ and noting in which profile this quantity saturates to a fixed non-zero value as one carries out this experiment placing detectors with multiple mean switching $\tau_{0}$, i.e., by switching on the detectors at different times. In particular, in Table \ref{tab:Obs} we have listed all our key observations regarding the radiative process of entangled detectors in the presence of a GW burst for both eternal and finite Gaussian switching.
    
\end{itemize}

  We would like to point out that the Wightman function \eqref{eq:greensFn-MGW-general-3} as obtained in our present manuscript corresponding to a specific geodesic trajectory is not different from the Wightman function corresponding to two static detectors, see \cite{Barman:2023aqk} where the authors investigate the entanglement harvesting condition. Therefore, one cannot distinguish whether the detectors are static or free-falling from the outcomes presented in the current work. It is to be noted that in both \cite{Barman:2023aqk} and the present manuscript, the Wightman functions are obtained considering terms $\mathcal{O}(\mathcal{A})$ in the field mode expansion, and then information about the detector trajectories are inserted. In \cite{Barman:2023aqk}, the trajectories were independent of $\mathcal{A}$ as the detectors were static. In contrast, in our present manuscript, the detector trajectories \eqref{eq:geodesic-trajectory} are also dependent on $\mathcal{A}$. However, that dependence does not influence the final expression of the Wightman function as compared to the static ones up to $\mathcal{O}(\mathcal{A})$ terms. We believe keeping higher-order terms in the Wightman function can help distinguish static and free-falling detector trajectories. For instance, in Appendix \ref{Appn:OA2-terms-WightmanFn} we have obtained the Wightman function corresponding to our specific geodesic trajectories \eqref{eq:geodesic-trajectory} keeping terms $\mathcal{O}(\mathcal{A}^2)$ and 
  {  it does depend on the specifics of the geodesic trajectory}. 
  In this regard, see the expression of Eq. \eqref{eq:WightmanFn-OA2-Trjctry}, which is dependent on the specific parameter $b$ characteristic of the geodesic trajectory \eqref{eq:geodesic-trajectory}. Thus, we conclude that by keeping $\mathcal{O}(\mathcal{A}^2)$ terms, one might be able to distinguish the specific detector trajectories in addition to identifying GW memory utilizing the radiative process of entangled quantum probes.

It has been known that the dynamics of entangled quantum probes can be affected by various system parameters such as the motion of the detectors, background curvature, thermal bath, spacetime dimensions, passing of gravitational wave, etc. In this regard, one can look into the introduction of the present manuscript for references. However, even if it depends on the passing of GW, whether it can also be affected by the memory in GW was not known previously. In this work, we considered a specific set-up comprising entangled quantum probes and discovered that it can identify GW memory backgrounds. For both eternal and Gaussian switching, we observed qualitative distinctions between the GW profiles with and without memory. The next immediate area to understand is how feasible these distinctions are in case one tries to measure them in an experimental setup. In this regard, we believe an interesting and practical way forward could be through the utilization of quantum communication and key distribution, see  \cite{Piveteau:2022gbr, Bruschi:2013sua, Bruschi:2014cma, Barzel:2022tbf}, where it is observed that quantum key distribution (QKD) protocols can efficiently capture the distortions in the photon wave packets caused due to the presence of curvature. An alternative practical direction involves examining the distortions in atomic electron transitions caused by the presence of GWs. This has been explored in hydrogen-like atoms, as discussed in \cite{Chen:2023wpp}, {and in the context of Rydberg atoms, as analyzed in \cite{Fischer:1994ed}.} These understandings can then be used to construct viable experimental set-ups in identifying the GW memory. 
We also note that studying the radiative process in asymptotically flat spacetime with generic GW waveforms from compact sources, such as those described by a Bondi-Sachs metric \cite{Madler:2016xju}, would give a more physically interesting picture. However, before addressing these scenarios, it is necessary to further develop quantum field theory in such non-trivial curved spacetimes.
We are currently working in these directions and wish to present some of our findings in a future communication.

%%%%%%%%%%%%%%%%%%%%%%%%%%%%%%%%%%%%%%%%%%%%%%%%%%%%%%%%%%%%%%%%%%%%%%%%%%%%%%%%%%%%%%%%%%%%%%%%%%%%%

\begin{acknowledgments}

We thank Sumanta Chakraborty for the initial idea for the project, and Bibhas Ranjan Majhi for comments on the manuscript. We would also like to thank Dawood Kothawala for the useful discussions and his comments on the manuscript. The authors would also like to acknowledge the anonymous referee
whose insightful comments helped in enhancing the quality of the paper. S.B. would like to thank the Science and Engineering Research Board (SERB), Government of India (GoI), for supporting this work through the National Post Doctoral Fellowship (N-PDF, File number: PDF/2022/000428). S.M. would like to acknowledge the Inspire Faculty Grant (DST/INSPIRE/04/2022/001332), DST, Government of India, and New Faculty Seed Grant (NFSG/PIL/2023/P3794), BITS Pilani, Pilani, for financial support. 

\end{acknowledgments}

%%%%%%%%%%%%%%%%%%%%%%%%%%%%%%%%%%%%%%%%%%%%%%%%%%%%%%%%%%%%%%%%%%%%%%%
\appendix

\begin{widetext}
\section{Calculation of $\mathcal{O}(\mathcal{A}^2)$ terms in the Wightman function} \label{Appn:OA2-terms-WightmanFn}

In this section of the Appendix, we obtain the Wightmann function keeping terms $\mathcal{O}(\mathcal{A}^2)$. First, the expansion of field mode \eqref{eq:modeFn-general} is considered upto $\mathcal{O}(\mathcal{A}^2)$ to get the Wightman function. Second, in that resulting expression the information of appropriate detector trajectories, which can depend on $\mathcal{A}$ as well, are inserted to obtain the final form of the Wightman function. One can obtain expression of the Wightman function keeping terms $\mathcal{O}(\mathcal{A}^2)$ in the field mode expansion as
\begin{eqnarray}\label{eq:WightmanFn-OA2}
G_{W}(\tilde{x},\tilde{x}') = G_{W}^{0}(\tilde{x},\tilde{x}') + G_{W}^{1}(\tilde{x},\tilde{x}') + G_{W}^{2}(\tilde{x},\tilde{x}')~,
\end{eqnarray}
where the expressions of $G_{W}^{0}(\tilde{x},\tilde{x}')$ and $G_{W}^{1}(\tilde{x},\tilde{x}')$ are given in Eq. \eqref{eq:greensFn-MGW-general}, and the expression of $G_{W}^{2}(\tilde{x},\tilde{x}')$ is given by
\begin{eqnarray}\label{eq:WightmanFn-OA2-form}
    G_{W}^{2}(\tilde{x},\tilde{x}') &=& -\frac{\mathcal{A}^2\,\left\{\Bar{g}(u)-\Bar{g}(u')\right\}^2}{4\pi^2\,\Delta u^{5}}\,\bigg[\frac{\Delta u^2}{\frac{\sigma_{M}}{\Delta u}+i\,\epsilon}-\frac{2\Delta u\,(\Delta x^2+\Delta y^2)}{\left(\frac{\sigma_{M}}{\Delta u}+i\,\epsilon\right)^2}+\frac{2\,(\Delta x^2-\Delta y^2)^2}{\left(\frac{\sigma_{M}}{\Delta u}+i\,\epsilon\right)^3}\bigg]~.
\end{eqnarray}
The expression of the above $G_{W}^{2}(\tilde{x},\tilde{x}')$ was obtained with the help of \cite{Barman:2023aqk} (see Appendix A of \cite{Barman:2023aqk}). Putting the information of detector trajectories from Eq. \eqref{eq:geodesic-trajectory} one can obtain the above Wightman function (keeping terms $\mathcal{O}(\mathcal{A}^2)$) connecting two different detector events as
\begin{eqnarray}\label{eq:WightmanFn-OA2-Trjctry}
    G_{W}(\tilde{x}_{1},\tilde{x}_{2}) &\simeq& -\frac{1}{4\pi^2}\, \frac{1}{\Delta \tau^2-d^2-i\,\epsilon\,\Delta \tau}-\frac{1}{4\pi^2}\, \frac{\mathcal{A}\, \left\{\Bar{g}(\tau_{1})-\Bar{g}(\tau_{2})\right\}}{(\Delta \tau^2-d^2-i\,\epsilon\,\Delta \tau)^2}\,\frac{d^2}{\Delta \tau}\nonumber\\
    ~&+& \frac{1}{4\pi^2}\, \frac{\mathcal{A}^2\, \left\{\Bar{g}(\tau_{1})-\Bar{g}(\tau_{2})\right\}^2}{(\Delta \tau^2-d^2-i\,\epsilon\,\Delta \tau)\,\Delta \tau^2}\,\bigg[1+\frac{b^2\,\Delta \tau^2+2\,d^2}{(\Delta \tau^2-d^2-i\,\epsilon\,\Delta \tau)}+\frac{2\,d^4-6\,d^2 b^2\Delta \tau^2+3\,b^4\Delta \tau^4}{(\Delta \tau^2-d^2-i\,\epsilon\,\Delta \tau)^2}\bigg]~.
\end{eqnarray}
\end{widetext}
From the above expression of the Wightman function, we observe that it is $b$ dependent, in contrast to the Wightman function of Eq. \eqref{eq:greensFn-MGW-general-3} where $\mathcal{O}(\mathcal{A})$ terms were considered. Therefore, one can conclude that by considering higher order of $\mathcal{A}$ terms in the Wightman function the explicit signatures of the detector trajectory can be identified. However, we would like to mention that contribution due to $\mathcal{O}(\mathcal{A}^2)$ terms are much smaller as compared to the $\mathcal{O}(\mathcal{A})$ terms, and we could distinguish the effect of GW memory even with terms $\mathcal{O}(\mathcal{A})$ which is a significant observation.

\color{black}

%\-lobibliographystyle{apsrev}
\bibliographystyle{utphys1.bst}

\bibliography{bibtexfile}

\end{document}